\documentclass[aps,preprint,nofootinbib,eqsecnum]{revtex4}%
\usepackage{amsfonts}
\usepackage{amsmath}
\usepackage{amssymb}
\usepackage{graphicx}%
\providecommand{\U}[1]{\protect\rule{.1in}{.1in}}
\providecommand{\U}[1]{\protect\rule{.1in}{.1in}}

\begin{document}

\leftline {USC-08/HEP-B5 \hfill ArXiv: 0811.2510 [hep-th]} {\vskip-1cm}

{\vskip2cm}

\begin{center}
{{\LARGE Geometry and Symmetry Structures in 2T Gravity}\footnote{This work
was partially supported by the US Department of Energy under grant number
DE-FG03-84ER40168.}}

{\vskip1cm}

\textbf{Itzhak Bars and Shih-Hung Chen}

{\vskip0.5cm}

\textsl{Department of Physics and Astronomy}

\textsl{University of Southern California,\ Los Angeles, CA 90089-2535 USA}

{\vskip1.5cm} \textbf{Abstract}
\end{center}

2T-gravity in $d+2$ dimensions predicts 1T General Relativity (GR) in $d$
dimensions, augmented with a local scale symmetry known as the Weyl symmetry
in 1T field theory. The emerging GR comes with a number of constraints,
particularly on scalar fields and their interactions in 1T field theory. These
constraints, detailed in this paper, are footprints of 2T-gravity and could be
a basis for testing 2T-physics. Some of the conceptually interesting
consequences of the \textquotedblleft accidental\textquotedblright\ Weyl
symmetry include that the gravitational constant emerges from vacuum values of
the dilaton and other Higgs-type scalars and that it changes after every
cosmic phase transition (inflation, grand unification, electroweak phase
transition, etc.). We show that this consequential Weyl symmetry in $d$
dimensions originates from coordinate reparametrization, not from scale
transformations, in the $d+2$ spacetime of 2T-gravity. To recognize this
structure we develop in detail the geometrical structures, curvatures,
symmetries, etc. of the $d+2$ spacetime which is restricted by a homothety
condition derived from the action of 2T-gravity. Observers that live in $d$
dimensions perceive GR and all degrees of freedom as shadows of their
counterparts in $d+2$ dimensions. Kaluza-Klein (KK) type modes are removed by
gauge symmetries and constraints that follow from the 2T-gravity action.
However some analogs to KK modes, which we call \textquotedblleft
prolongations\textquotedblright\ of the shadows into the higher dimensions,
remain but they are completely determined, up to gauge freedom, by the shadows
in $d$ dimensions.

\newpage

\tableofcontents
\newpage

\section{Introduction}

Two-Time Gravity \cite{2Tgravity} in $d+2$ dimensions has successfully
reproduced the usual one time General Relativity as a shadow in $\left(
d-1\right)  +1$ \ dimensions. Taken together with similar recent results for
the Standard Model \cite{2tstandardM} and $\mathcal{N}=1,2,4$ supersymmetric
2T field theory \cite{susy2tN1}\cite{susy2tN24}, these 2T theories correctly
describe $3+1$ dimensional Nature directly in $4+2$ dimensions. The
phenomenologically successful theories now have counterparts in $4+2$
dimensions, thus providing a new perspective on the significance of space and
time and lending a new outlook on unification of 1T-physics theories.

Briefly, the relation between the $4+2$ and $3+1$ theory is as follows. After
gauge fixing and solving some kinematic equations of motion, the $4+2$ field
theory yields various \textquotedblleft shadows\textquotedblright\ in $3+1$
dimensions. The \textquotedblleft conformal shadow\textquotedblright\ of the
$4+2$ theories coincide with the standard familiar theories, except for some
additional new constraints. These new constraints on 1T field theory - in
particular on scalar fields and their interactions - are consistent with
everything we know so far. Potentially there are measurable phenomenological
consequences of these new restrictions within the conformal shadow that could
distinguish 2T-physics from other approaches, as explained at the end of
section (\ref{review}).

In addition, a main novelty in 2T-physics is that this formalism produces many
1T-physics shadows from the same parent theory. The conformal shadow mentioned
in the previous paragraph is only one of many. The other shadows provide 1T
field theories that are dual to the familiar ones and these may be turned into
computational tools for extracting non-perturbative physics. The shadows give
different perspectives of the $4+2$ theory as viewed by observers that are
stuck in $3+1$ dimensions. The different embedding of $3+1$ dimensions into
$4+2$ dimensions contain moduli that appear in $3+1$ dimensions as parameters
of the 1T shadow theory, such as mass, curvature or interaction with
backgrounds, which offer different glimpses of the higher dimensions.
Dualities transform shadows with different $3+1$ geometries or different
values of the parameters. The shadows and dualities are most easily understood
in the worldline formulation of 2T-physics\footnote{For examples of $\left(
d-1\right)  +1$ shadows that emerge from flat $d+2$ spacetime, see tables
I,II,III in \cite{emergentfieldth1}. \label{examples}}. While the
investigation of dualities in the 1T field theory formalism are ongoing
\cite{massiveGauge}, some of the simpler cases have been reported in
\cite{emergentfieldth1} for scalar fields and in \cite{emergentfieldth2} for
Dirac and Yang-Mills fields.

Through the dualities, and through hidden symmetries related to the higher
spacetime, the parent theory in $d+2$ dimensions provides a new kind of
unification of various 1T-physics field theories.

In this paper we will concentrate exclusively on the conformal shadow of 2T
gravity in $d+2$ dimensions in order to clarify further its geometrical and
symmetry properties. Specifically, we will investigate not only the shadow in
$\left(  d-1\right)  +1$ dimensions but also its prolongation into $d+2$
dimensions. By this we mean that there are Riemann curvature components
$R_{~NPQ}^{M}$ and other geometrical fields that are non-vanishing not only in
the shadow in $\left(  d-1\right)  +1$ dimensions but also in $d+2$
dimensions. We will show that all such non-vanishing components of the
prolongation of the shadow are actually fully determined, up to gauge freedom,
by the fields within the shadow in $\left(  d-1\right)  +1$ dimensions.

Concentrating only on the shadow with an effective action principle in
$\left(  d-1\right)  +1$ is self consistent as shown in \cite{2Tgravity}.
However, the extension of the shadow into the higher spacetime is likely to be
important for discussing the dualities among shadows as well as for grasping
the higher $d+2$ dimensional properties of the underlying theory.

Another important property of the conformal shadow for gravity in $\left(
d-1\right)  +1$ is that General Relativity comes with a local rescaling Weyl
symmetry \cite{zumino}-\cite{jackiw}, along with a dilaton that compensates
for the local rescaling of the metric. This is one of the important
restrictions imposed on 1T-physics by 2T-physics, as reported in
\cite{2Tgravity}. The physical effect of this is that the gravitational
constant is not a parameter but emerges in 1T-physics from the vacuum value of
the dilaton\footnote{The massless Goldstone boson that emerges from the
spontaneous breaking of scale symmetry \cite{zumino}-\cite{dilaton7}, i.e. the
fluctuation of the dilaton around its vacuum value, is eliminated by a Weyl
gauge choice in our theory, so it does not generate any long range forces that
could compete with the long range effects of gravity \cite{sundrum}.}.

A further property associated with the Weyl symmetry is that every scalar
field in 1T-physics beyond the dilaton (such as inflaton, Higgs, etc.) must be
a conformal scalar that has a special fixed dimensionless coupling to the
curvature scalar $R.$ The physical effect of this is that the gravitational
constant changes as a function of cosmic time after every phase transition in
the universe (inflation, grand unification, etc.). In this paper we will
clarify the origin of this important accidental local scale symmetry in the
conformal shadow. It will be shown that it emerges as a remnant from symmetry
under coordinate transformation (not scale transformations) of the higher
dimensional 2T-Gravity.

In section (\ref{review}) we will briefly review the basic setup of
2T-Gravity, display its reduction to ordinary 1T General Relativity augmented
with the local Weyl symmetry, and explain the physically significant
constraints that this structure puts on 1T field theory coupled to gravity.
The rest of the paper develops the technical aspects of the geometry and
symmetries to explain in detail how the reduced 1T theory of section
(\ref{review}) is recovered from 2T-gravity. In section (\ref{kinematicsd+2})
we discuss the kinematics of 2T curved spacetime in $d+2$ dimensions. This
involves solving the kinematical equations of motion that follow from the
2T-gravity action and working out the general consequences that the geometry
of the 2T spacetime is restricted by a homothety condition on the metric. In
section (\ref{eoms}) the dynamical and kinematical equations are discussed and
their relation to an Sp$\left(  2,R\right)  $ gauge symmetry of an underlying
worldline particle theory is explained. In section (\ref{shadow}) we show how
spacetime in $\left(  d-1\right)  +1$ dimensions is embedded in spacetime in
$\left(  d+2\right)  $ dimensions by making gauge choices and solving the
kinematic equations. This leads to an explanation of the origin of the local
scaling symmetry in General Relativity in $\left(  d-1\right)  +1$ dimensions
known as the Weyl symmetry. It will be shown that it originates from general
coordinate transformations, not from local rescalings, in $d+2$ dimensions. In
section (\ref{curve}) we calculate the components of the Riemann tensors and
of the SO$\left(  d,2\right)  $ \textquotedblleft Lorentz\textquotedblright%
\ curvature in tangent space, that describe the geometry of the prolongation
of the conformal shadow into the higher dimensions. In section (\ref{dyn}) we
discuss in more detail the emerging 1T dynamical equations of motion of both
the shadow fields in $d$ dimensions and their prolongations to higher
dimensions and show that, up to gauge freedom, the prolongations are
completely determined by the shadow fields in $d$ dimensions. This leads to
one of our main conclusions, that the shadow fields themselves are determined
self consistently by the action only within the shadow in $d$ dimensions,
\textit{independently of the prolongations}, which was one of the goals in our investigation.

\section{Constraints in 1T field theory induced by 2T-Gravity}

\label{review}

In this section we briefly review 2T-gravity to explain the constraints that
it induces in 1T field theory, particularly involving scalar fields. We will
see that the gravitational constant emerges from the vacuum values of the
scalars and that it appears in several places in the action of 1T field
theory. The structure of scalars that emerges in 1T field theory, shown in
Eq.(\ref{nonlinear2}) and related discussion, is consistent with current
observations but this structure could be one of the future tests for the
predictions of 2T-Physics.

2T-Gravity, without any matter, includes three fields which we call the
gravity triplet: the metric $G_{MN},$ the dilaton $\Omega$ and another scalar
field $W,$ all in $d+2$ dimensions $X^{M\ }$. The action for pure 2T gravity
is \cite{2Tgravity}
\begin{equation}
S^{0}=\gamma\int d^{d+2}X\sqrt{G}\left\{
\begin{array}
[c]{c}%
\delta\left(  W\right)  \left[  \Omega^{2}R\left(  G\right)  +\frac{1}%
{2a}\partial\Omega\cdot\partial\Omega-V\left(  \Omega\right)  \right] \\
+\delta^{\prime}\left(  W\right)  \left[  \Omega^{2}\left(  4-\nabla
^{2}W\right)  +\partial W\cdot\partial\Omega^{2}\right]
\end{array}
\right\}  . \label{action}%
\end{equation}
Here $R\left(  G\right)  $ is the Riemann curvature scalar, $a$ is the special
constant%
\begin{equation}
a\equiv\frac{d-2}{8\left(  d-1\right)  }, \label{a value}%
\end{equation}
while the potential $V$ can only have the form $V\left(  \Omega\right)
=\lambda\Omega^{\frac{2d}{d-2}}$ with a dimensionless coupling $\lambda.$ The
overall constant $\gamma$ can be absorbed away by rescaling the fields, but is
used for convenience to normalize the 1T shadow action that emerges in two
lower dimensions. The action with this structure of kinetic terms, value of
$a$ and form of $V$ is unique under certain local gauge symmetries discussed
in \cite{2Tgravity}.

The unusual features of this action as a field theory include the delta
function $\delta\left(  W\right)  $ and its derivative $\delta^{\prime}\left(
W\right)  .$ All fields are varied freely to derive equations of motion or to
verify symmetries. The variations contain terms proportional to $\delta\left(
W\right)  ,\delta^{\prime}\left(  W\right)  ,\delta^{\prime\prime}\left(
W\right)  $ where more derivatives on $\delta\left(  W\right)  $ or
$\delta^{\prime}\left(  W\right)  $ emerge from integration by parts and the
chain rule $\partial_{M}\delta\left(  W\right)  =\delta^{\prime}\left(
W\right)  \partial_{M}W,$ etc. The coefficients of $\delta\left(  W\right)
,\delta^{\prime}\left(  W\right)  ,\delta^{\prime\prime}\left(  W\right)  $
for each general variation $\delta G^{MN},\delta\Omega,\delta W$ give three
equations of motion for each field. We will discuss some of the equations
motion later. There are remarkable consistencies between these equations all
due to the noteworthy symmetries of this action. This symmetry in field theory
captures the essentials of an underlying Sp$\left(  2,R\right)  $ symmetry
(see section (\ref{eoms})) that makes position and momentum $X^{M}\left(
\tau\right)  ,P_{M}\left(  \tau\right)  $ indistinguishable at every instant
$\tau$ at the level of a worldline formulation of a particle in the presence
of gravity \cite{2Tgravity}.

Part of the gauge symmetry can be used to fix $W\left(  X\right)  $ to any
function of $X^{M}$ that can vanish in some region of spacetime $X^{M}$. To
understand the role of $W$ the reader is reminded that, in 2T field theory in
\textit{flat space,} $W$ is replaced by a fixed function $W_{flat}$%
=$X^{M}X^{N}\eta_{MN}$ where $\eta_{MN}$ is the SO$\left(  d,2\right)  $
invariant flat metric. When 2T field theory in flat $d+2$ dimensions is
reduced to shadows in $\left(  d-1\right)  +1$ dimensions, then, in the
conformal shadow, the SO$\left(  d,2\right)  $ symmetry of $W$ becomes the
conformal symmetry of 1T field theory in Minkowski space\footnote{In other 1T
shadows$^{\text{\ref{examples}}}$ this SO$(d,2)$ of the flat 2T theory is
still a symmetry that is usually hidden and often not noticed in 1T-physics
before discovering it through a shadow of 2T-physics.}.

The symmetries of the action (\ref{action}) do not allow a gravitational
constant in $d+2$ dimensions, however Newton's constant emerges in the shadow
in $d$ dimensions from the vacuum value of the dilaton $\langle\Omega\rangle$
when the equations of motion are used to reduce this theory to the conformal
shadow. The conformal shadow action derived from (\ref{action}) in
\cite{2Tgravity}) (see section (\ref{dyn}) for justification) has the familiar
form of a conformal scalar
\begin{equation}
S_{shadow}^{0}=\int d^{d}x\sqrt{-g}\left(  \frac{1}{2a}g^{\mu\nu}\partial
_{\mu}\phi\partial_{\nu}\phi+R\phi^{2}-V\left(  \phi\right)  \right)  ,
\label{confScalar}%
\end{equation}
where $g_{\mu\nu}\left(  x\right)  $ is the metric in $d$ dimensions,
$R\left(  g\right)  $ is its Riemann curvature scalar and $a$ has the special
value in Eq.(\ref{a value}). The relation of $g_{\mu\nu}\left(  x\right)  $ to
$G_{MN}\left(  X\right)  $ and of $\phi\left(  x\right)  $ to $\Omega\left(
X\right)  $ will be displayed below. Suffice it for now to say that $\left(
g_{\mu\nu},\phi\right)  $ are the shadows of the higher dimensional fields as
seen by observers living in $d$ dimensions. There are no Kaluza-Klein (KK)
type physical degrees of freedom, as those are removed by the gauge symmetries
of 2T-physics. But there are some analogs to KK modes, which we call
\textquotedblleft prolongations\textquotedblright\ of the shadow into the
higher dimensions, determined by the shadow fields $\left(  g_{\mu\nu}%
,\phi\right)  $ as will be discussed later in this paper.

In the conformal shadow in Eq.(\ref{confScalar}) there is an accidental Weyl
symmetry that plays multiple important roles. Due to the special value of $a,$
Eq.(\ref{confScalar}) is the well known action of a conformal scalar that has
a Weyl symmetry $S_{shadow}^{0}\left(  g^{\prime},\phi^{\prime}\right)
=S_{shadow}^{0}\left(  g,\phi\right)  $ under local rescalings
\cite{WeylSymmetry} $g_{\mu\nu}^{\prime}\left(  x\right)  =e^{2\lambda\left(
x\right)  }g_{\mu\nu}\left(  x\right)  $,\ $\phi^{\prime}\left(  x\right)
=e^{-\frac{d-2}{2}\lambda\left(  x\right)  }\phi\left(  x\right)  $ with an
arbitrary $\lambda\left(  x\right)  .$ The original action in $d+2$ dimensions
(\ref{action}) does not have a Weyl symmetry, so the symmetry in the conformal
shadow appears to be \textquotedblleft accidental\textquotedblright. Later in
this paper it will be explained how this symmetry originates in the coordinate
transformations (not in scale transformations) in higher dimensions. Using
this local symmetry, $\phi\left(  x\right)  $ can be gauge fixed to a constant
$\phi_{0}$, so that the action (\ref{confScalar}) becomes precisely pure
General Relativity in $d$ dimensions
\begin{equation}
S_{shadow}^{0,fixed}=\int d^{d}x\sqrt{-g}\left(  \frac{R\left(  g\right)
}{2\kappa_{d}^{2}}-\frac{\Lambda_{d}}{2\kappa_{d}^{2}}\right)  ,\;\text{with }%
\begin{array}
[c]{l}%
\frac{1}{2\kappa_{d}^{2}}\equiv\phi_{0}^{2},\\
\Lambda_{d}\equiv\lambda\left(  \phi_{0}\right)  ^{\frac{4}{d-2}}.
\end{array}
\label{S0shad}%
\end{equation}

Note that according to the sign of the kinetic term in Eq.(\ref{confScalar}),
the field $\phi\left(  x\right)  $ has negative norm, but this sign is
required in order to obtain a positive gavitational constant while being
consistent with the Weyl symmetry. Of course, by having the Weyl symmetry, the
negative norm ghost, which is also a Goldstone boson of scale transformations,
is removed from the physical spectrum. This nice feature is a consequence of
the symmetries of the higher dimensional 2T-gravity theory.

When matter is included the Weyl gauge can be chosen in various other ways
(see below), and then one finds more physical effects of the dilaton beyond
its footprints in the form of the gravitational constant in $d$ dimensions
$\left(  2\kappa_{d}^{2}\right)  ^{-1}$ and an undetermined cosmological
constant $\Lambda_{d}$ ($\lambda$ has any sign or magnitude).

We now outline the coupling of the gravity triplet $\left(  W,\Omega
,G^{MN}\right)  $ to matter fields of the type Klein-Gordon $S_{i}\left(
X\right)  $, Dirac $\Psi\left(  X\right)  $ and Yang-Mills $A_{M}\left(
X\right)  $ \cite{2Tgravity}. In \textit{flat} 2T field theory these fields
must have the following engineering dimensions \cite{2tstandardM}
\begin{equation}
\dim\left(  X^{M}\right)  =1,\;\dim\left(  S_{i}\right)  =-\frac{d-2}%
{2},\;\dim\left(  \Psi\right)  =-\frac{d}{2},\;\dim\left(  A_{M}\right)  =-1
\label{dims}%
\end{equation}
The general 2T field theory of these fields in flat space in $d+2$ dimensions
was given in \cite{2tstandardM}. The matter part of the theory in curved space
follows from the flat theory in \cite{2tstandardM} by making the substitutions
indicated in Table-1 \cite{2Tgravity}
\begin{gather*}
\;\;\;\;\;%
\begin{tabular}
[c]{|l|l|l|}\hline
Quantity & Flat & Curved\\\hline
metric & $\eta^{MN}$ & $G^{MN}\left(  X\right)  $\\
volume element~ & $\left(  d^{d+2}X\right)  \delta\left(  X^{2}\right)  \;\;$
& $\left(  d^{d+2}X\right)  \sqrt{G}\delta\left(  W\left(  X\right)  \right)
$\\
explicit $X$ & $X^{M}$ & $V^{M}\left(  X\right)  =\frac{1}{2}G^{MN}%
\partial_{N}W$\\
gamma matrix, vielbein~ & $\gamma_{M}$ & $E_{M}^{a}\left(  X\right)
\gamma_{a}$\\
spin connection~ & $\gamma^{M}\partial_{M}\Psi~$ & $E^{Mc}\gamma_{c}\left(
\partial_{M}+\frac{1}{4}\gamma_{ab}~\omega_{M}^{ab}\left(  X\right)  \right)
\Psi$\\
Yang-Mills & specialize $\eta^{MN}$ & $\Omega^{\frac{2\left(  d-4\right)
}{d-2}}~Tr\left(  -\frac{1}{4}F_{MN}F_{KL}\right)  G^{MK}G^{NL}$\\
Yukawa & specialize $X^{M}\gamma_{M}\;$ & $\Omega^{-\frac{d-4}{d-2}}%
~V^{M}\left(  g_{i}\bar{\Psi}^{L}\gamma_{M}\Psi^{R}S_{i}+h.c.\right)  $\\
$\left\{
\begin{array}
[c]{l}%
\text{real scalar fields }S_{i},\Omega\\
\text{extra}~\frac{-1}{a}~\text{for dilaton~~~ }%
\end{array}
\right\}  $ & $\left\{
\begin{array}
[c]{c}%
\text{complex }\\
\varphi=\frac{S_{1}+iS_{2}}{\sqrt{2}}%
\end{array}
\right\}  \;$ & $\left\{
\begin{array}
[c]{l}%
G^{MN}\left(  \frac{1}{2a}\partial_{M}\Omega\partial_{N}\Omega-\frac{1}{2}%
\sum_{i}\partial_{M}S_{i}\partial_{N}S_{i}\right) \\
+\left(  \Omega^{2}-a\sum_{i}S_{i}^{2}\right)  R\left(  G\right)  -V\left(
\Omega,S_{i}\right)
\end{array}
\right\}  $\\
$\delta^{\prime}\left(  W\right)  $ term, scalars only & $W_{flat}=X^{2}$ &
$\left\{
\begin{array}
[c]{c}%
\left(  \Omega^{2}-a\sum_{i}S_{i}^{2}\right)  \left(  4-\nabla^{2}W\right) \\
+\partial W\cdot\partial\left(  \Omega^{2}-a\sum_{i}S_{i}^{2}\right)
\end{array}
\right\}  \delta^{\prime}\left(  W\right)  $\\\hline
\end{tabular}
\ \ \ \ \ \ \ \ \ \\
\text{Table-1. Matter }S_{i},\Psi,A_{M}\text{ in interaction with the gravity
triplet }\left(  W,\Omega,G^{MN}\right)  \text{. }%
\end{gather*}
The dilaton $\Omega$ couples to Yang-Mills fields with factor $\Omega
^{\frac{2\left(  d-4\right)  }{d-2}}$ and Yukawa terms with factor
$\Omega^{-\frac{d-4}{d-2}}$ as in Table-1. This coupling of $\Omega$ is
dictated by the symmetries of the theory consistently with the dimensions in
Eq.(\ref{dims}). When $d+2=6,$ these factors become $1,$ so this coupling of
the dilaton disappears for this special case. An important property of
$V\left(  \Omega,S_{i}\right)  $ related again to the dimensions (\ref{dims})
and symmetries, is that it must have the homogeneity property\footnote{Then
$V\left(  \Omega,S_{i}\right)  $ has dimension $-d$ under the scaling of
scalars $\left(  \Omega,S_{i}\right)  \rightarrow e^{-\frac{d-2}{2}\lambda
}\left(  \Omega,S_{i}\right)  $, that is $V\rightarrow e^{-d\lambda}V.$}
\begin{equation}
V\left(  t\Omega,tS_{i}\right)  =t^{\frac{2d}{d-2}}V\left(  \Omega
,S_{i}\right)  . \label{Vt}%
\end{equation}
A general function with this property may be written in the form $V\left(
\Omega,S_{i}\right)  =\Omega^{\frac{2d}{d-2}}f\left(  S_{i}/\Omega\right)  ,$
where $f\left(  \sigma_{i}\right)  $ is an arbitrary function of the scale
invariant variables $\sigma_{i}=S_{i}/\Omega.$

We emphasize an important property of the scalars $S_{i}$ (including the Higgs
field in the Standard Model). The symmetries require that, except for an
overall normalization for each scalar, the \textit{quadratic} part of the
Lagrangian for any scalar $S_{i}\left(  X\right)  $ must have exactly the same
structure as the one for the dilaton field $\Omega$ in the pure gravity action
Eq.(\ref{action}). This structure is included for scalars $\Omega,S_{i}$ in
Table-1, where the sign and structure of the curvature term relative to the
kinetic term is fixed by the constant $a$. Furthermore, the symmetry requires
also a $\delta^{\prime}\left(  W\right)  $ term for the quadratic term in the
scalars $\Omega,S_{i}$ as shown in the table\footnote{In flat space the
$\delta^{\prime}\left(  X^{2}\right)  $ term can be rewritten as a
$\delta\left(  X^{2}\right)  $ term that modifies the naive kinetic term. As
indicated in the Table, for 2T field theory in flat spacetime the function $W$
is replaced by $W_{flat}=X^{2}.$ Then it can be verified that for each scalar
$\Omega,S_{i}$ the kinetic terms $\left(  \frac{1}{2a}\partial\Omega
\cdot\partial\Omega-\frac{1}{2}\partial S_{i}\cdot\partial S_{i}\right)  $
that are multiplied with $\delta\left(  X^{2}\right)  $ combine with the
$\delta^{\prime}\left(  X^{2}\right)  $ terms in Table-1 to become simply
$\delta\left(  X^{2}\right)  \left(  \frac{1}{2a}\Omega\partial^{2}%
\Omega-\frac{1}{2}S_{i}\partial^{2}S_{i}\right)  $ after dropping a total
derivative, thus avoiding any $\delta^{\prime}\left(  X^{2}\right)  $ terms,
as in the general 2T-field theory in flat space \cite{2tstandardM}.}.

The overall sign and magnitude of the normalization for the kinetic term
($\frac{-1}{2}$) of a real scalar $\frac{-1}{2}G^{MN}\partial_{M}S_{i}%
\partial_{N}S_{i}$ is fixed by the requirements of unitarity (no negative
norms) and conventional definition of norm. For the dilaton the norm differs
by an overall $\frac{-1}{a};$ the magnitude can be changed by rescaling the
field $\Omega$ so it is not significant, but the sign is significant (negative
norm) and is needed to produce a positive Newton constant from the vacuum
values of the scalars in the shadow 1T theory as explained above. This
negative norm ghost is harmless since it is removable in the shadow by using
the leftover Weyl symmetry arising from coordinate transformations.

The form of the shadow action with only scalar matter fields was derived in
\cite{2Tgravity} (see section (\ref{dyn}) for justification). It has the form
of conformal scalars coupled to gravity
\begin{equation}
S_{shadow}\left(  g,\phi,s_{i}\right)  =\int d^{d}x\sqrt{-g}\left(
\begin{array}
[c]{c}%
\frac{1}{2a}g^{\mu\nu}\partial_{\mu}\phi\partial_{\nu}\phi-\frac{1}{2}%
g^{\mu\nu}\partial_{\mu}s_{i}\partial_{\nu}s_{i}\\
+\left(  \phi^{2}-as_{i}^{2}\right)  R-V\left(  \phi,s_{i}\right)
\end{array}
\right)  . \label{S0shad2}%
\end{equation}
where a sum over $i$ is implied. The equations of motion that follow from this
action include
\begin{equation}
R_{\mu\nu}\left(  g\right)  -\frac{1}{2}g_{\mu\nu}R\left(  g\right)
=T_{\mu\nu}\left(  \phi,s_{i}\right)  , \label{Rmn}%
\end{equation}
with the energy momentum tensor $T_{\mu\nu}$ given by%
\begin{equation}
T_{\mu\nu}=\frac{1}{\left(  \phi^{2}-as_{i}^{2}\right)  }\left[
\begin{array}
[c]{c}%
\left(  -\frac{1}{2a}\partial_{\mu}\phi\partial_{\nu}\phi+\frac{1}{2}%
\partial_{\mu}s_{i}\partial_{\nu}s_{i}\right)  -\left(  g_{\mu\nu}\nabla
^{2}-\nabla_{\mu}\partial_{\nu}\right)  \left(  \phi^{2}-as_{i}^{2}\right) \\
+\frac{1}{2}g_{\mu\nu}\left(  \frac{1}{2a}\partial\phi\cdot\partial\phi
-\frac{1}{2}\partial s_{i}\cdot\partial s_{i}-V\left(  \phi,s_{i}\right)
\right)
\end{array}
\right]  . \label{Tmunu}%
\end{equation}
The relation of the shadow $\left(  g_{\mu\nu},\phi,s_{i}\right)  $ to
$\left(  G_{MN},\Omega,S_{i}\right)  $ and their prolongations will be given
below. If there are $N$ real scalars $s_{i}$ in addition to the dilaton $\phi
$, then the kinetic and curvature terms have an automatic global symmetry
SO$\left(  N,1\right)  ,$ with a SO$\left(  N,1\right)  $ diagonal metric
$\left(  -1/a,1,1,\cdots,1\right)  $ as seen in the expression $\left(
\phi^{2}-as_{i}^{2}\right)  R$ and the kinetic term. This symmetry could be
explicitly broken by the potential $V\left(  \phi,s_{i}\right)  $ which is
arbitrary except for the homogeneity condition in Eq.(\ref{Vt})\footnote{Of
course, in a complete model of fundamental interactions, various Yang-Mills
gauge symmetries also put constraints on the structure of the potential
$V\left(  \Omega,S_{i}\right)  $ in addition to the homogeneity condition
(\ref{Vt}). \label{symmV}}. The vacuum is determined by the properties of
$V\left(  \phi,s_{i}\right)  $ and therefore the gravitational constant and
the cosmological constants are now functions of the vacuum values of all the
scalars
\begin{equation}
\frac{1}{2\kappa_{d}^{2}}=\langle\phi^{2}-as_{i}^{2}\rangle,\;\;\frac
{\Lambda_{d}}{2\kappa_{d}^{2}}=V\left(  \langle\phi\rangle,\langle
s_{i}\rangle\right)  , \label{scales}%
\end{equation}
generalizing Eq.(\ref{S0shad}). These gravitational and cosmological
\textquotedblleft constants\textquotedblright\ $\left(  \kappa_{d},\Lambda
_{d}\right)  $ induced by the various fundamental scalars $\left(
\Omega,S_{i}\right)  $ are not really constants since they must change after
every cosmic phase transition of the Universe as a whole (inflation, grand
unification, electroweak phase transition, etc.) as the various vacuum
expectations values $\langle\phi\rangle,\langle s_{i}\rangle$ turn on at
critical values of cosmic temperature or cosmic time. The cosmological
implications of this are under study \cite{2tcosmology}.

An important fact again is the presence of the accidental Weyl symmetry in the
action (\ref{S0shad2}), $S_{shadow}\left(  g^{\prime},\phi^{\prime}%
,s_{i}^{\prime}\right)  =S_{shadow}\left(  g,\phi,s_{i}\right)  ,$ under local
rescalings with an arbitrary \textit{local} gauge parameter $\lambda\left(
x\right)  $
\begin{equation}
g_{\mu\nu}^{\prime}=e^{2\lambda}g_{\mu\nu},\ \phi^{\prime}=e^{-\frac{d-2}%
{2}\lambda}\phi,\;s_{i}^{\prime}=e^{-\frac{d-2}{2}\lambda}\phi.
\label{rescale}%
\end{equation}
This symmetry persists in the shadow action with additional matter fields when
the fermions and gauge fields are included in the 2T action according to
Table-1. The Weyl symmetry can be used to remove the dilatonic Goldstone boson
(now a mixture of many fields $\phi,s_{i}$). The remaining physical scalar
fields, \textit{after} the phase transitions that produce the gravitational
constant $\left(  2\kappa_{d}^{2}\right)  ^{-1}$, can be neatly described by
fixing the Weyl gauge so that the dilaton $\phi$ gets determined by the other
scalars as follows
\begin{equation}
\phi\left(  x\right)  =\pm\left(  as_{i}^{2}\left(  x\right)  +\frac
{1}{2\kappa_{d}^{2}}\right)  ^{1/2}. \label{nonlinear}%
\end{equation}
This gauge choice reduces the curvature term in Eq.(\ref{S0shad2}) to simply
$R\left(  g\right)  /\left(  2\kappa_{d}^{2}\right)  ,$ thus conveniently
describing gravity after the phase transition in the Einstein frame. However,
while this gauge choice is convenient to describe gravity in the traditional
setting, the gravitational constant enters in a few other places in the action
as described below. In particular, the kinetic term of the scalars in the
shadow action (\ref{S0shad2}) turns into a nonlinear sigma model for the group
SO$\left(  N,1\right)  $ (see Eq.(\ref{nonlinear2})). The scale of the
non-linearity in the sigma model is determined by the gravitational constant
$\left(  2\kappa_{d}^{2}\right)  ^{-1}$ as in Eq.(\ref{nonlinear}).

Taking advantage of the homogeneity of the potential, and using $t=\left(
\phi^{2}-as_{i}^{2}\right)  ^{-1/2}$ in Eq.(\ref{Vt}) $V$ can be written in
the form%
\begin{equation}
V\left(  \phi,s_{i}\right)  =\left(  \phi^{2}-as_{i}^{2}\right)  ^{\frac
{d}{d-2}}\times V\left(  \frac{\phi}{\sqrt{\phi^{2}-as_{i}^{2}}},\frac{s_{i}%
}{\sqrt{\phi^{2}-as_{i}^{2}}}\right)  \equiv v\left(  \sigma_{i}\right)  .
\end{equation}
In the Weyl gauge (\ref{nonlinear}), after replacing the overall coefficient
by the constant $\left(  \phi^{2}-as_{i}^{2}\right)  ^{\frac{d}{d-2}}=\left(
2\kappa_{d}^{2}\right)  ^{-\frac{d}{d-2}}$ , and absorbing it into the
definition of $v\left(  \sigma_{i}\right)  ,$ we see that the leftover
$v\left(  \sigma_{i}\right)  $ is an \textit{arbitrary} function of the $N$
variables $\sigma_{i}=s_{i}/\phi$ that are invariants under the local scale
transformations of Eq.(\ref{rescale}). Of course at this fixed gauge there are
now some scales in the theory, namely the gravitational scale $\kappa_{d}$ and
the other scales $\langle s_{i}/\phi\rangle=\langle\sigma_{i}\rangle$
generated by phase transitions that follow from the properties of $v\left(
\sigma_{i}\right)  $.

The kinetic term of the scalars in the non-linear sigma model can also be
written in terms of the $\sigma_{i}$ or the $s_{i}.$ For example, if we
parametrize the $N$ fields as $s_{i}=sn_{i}$ where $n_{i}\left(  x\right)  $
is a unit vector $\sum_{i}n_{i}n_{i}=1$ and $s\left(  x\right)  $ is the
magnitude of the SO$\left(  N\right)  $ vector $s_{i}\left(  x\right)  $, then
we can write $\sum_{i}s_{i}^{2}=s^{2}$ so that $\phi$ in Eq.(\ref{nonlinear})
becomes a function of a single field $s\left(  x\right)  $
\begin{equation}
\phi=\pm\left(  2\kappa_{d}^{2}\right)  ^{-1/2}\sqrt{1+2a\kappa_{d}^{2}s^{2}}%
\end{equation}
Replacing these forms in the action (\ref{S0shad2}) we obtain an ordinary
looking General Relativity (after the phase transitions) coupled to an
arbitrary$^{\text{\ref{symmV}}}$ potential $v\left(  s,n_{i}\right)  $ with a
non-ordinary kinetic energy term for scalars where the gravity scale
$\kappa_{d}$ appears non-trivially%
\begin{equation}
S_{shadow}^{fixed}\left(  g,s,n_{i}\right)  =\int d^{d}x\sqrt{-g}\left\{
\begin{array}
[c]{c}%
\frac{1}{2\kappa_{d}^{2}}R\left(  g\right)  -v\left(  s,n_{i}\right) \\
-\frac{1}{2}g^{\mu\nu}\left[  \frac{\partial_{\mu}s\partial_{\nu}s}%
{1+2a\kappa_{d}^{2}s^{2}}+s^{2}\sum_{i}\partial_{\mu}n_{i}\partial_{\nu}%
n_{i}\right]
\end{array}
\right\}  . \label{nonlinear2}%
\end{equation}
The last term $\sum_{i}\partial n_{i}\cdot\partial n_{i},$ with $\vec{n}%
\cdot\vec{n}=1,$ is a non-linear SO$\left(  N\right)  $ sigma model, while
taken together as a whole the scalar kinetic terms form a non-linear
SO$\left(  N,1\right)  $ sigma model coupled to gravity. The scale of
non-linearity of the SO$\left(  N,1\right)  $ sigma model is also
\textit{determined uniquely} by the gravity scale $\kappa_{d}$ and the
constant $a$ given in Eq.(\ref{a value}). That the gravitational constant
$\left(  2\kappa_{d}^{2}\right)  ^{-1}$ should appear in this way in 1T field
theory, in addition to the traditional term $R\left(  g\right)  /2\kappa
_{d}^{2}$, is a prediction of the symmetries of 2T-gravity.

Additional places in 1T field theory action where $\left(  2\kappa_{d}%
^{2}\right)  ^{-1}$ appears as a consequence of 2T-gravity include the dilaton
factor $\phi^{\frac{2\left(  d-4\right)  }{d-2}}$ for Yang-Mills kinetic
terms, and the dilaton factor $\phi^{-\frac{d-4}{d-2}}$ for Yukawa terms,
which come from the similar terms in 2T field theory as shown in Table-1. In
these expressions $\phi=\pm\left(  2\kappa_{d}^{2}\right)  ^{-1/2}%
\sqrt{1+2a\kappa_{d}^{2}s^{2}}$ as above. Evidently when $d=4$ these factors
disappear, but they could play a physical role in a theory with $d$ less than
or larger than $4,$ thus providing additional signals of 2T-physics.

The $\sigma_{i}$ or $s_{i}$ can also be parametrized in other convenient ways
to take advantage of both the SO$\left(  N,1\right)  $ symmetry of the kinetic
term and of possible other symmetries$^{\text{\ref{symmV}}}$ of the potential
term $V\left(  \Omega,S_{i}\right)  .$ Presumably the symmetries of the
potential, indeed of the full theory, include at the very least the SU$\left(
3\right)  \times$SU$\left(  2\right)  \times$U$\left(  1\right)  $ symmetry of
the Standard Model, which is possibly embedded in a larger grand unified
symmetry group. In this context $N$ is the total number of all the real
scalars in the theory besides the dilaton. In the physical applications of
these concepts in a complete theory, one would be advised to take advantage of
the flavor/color or grand unified symmetry structures in choosing the most
convenient parametrization of the SO$\left(  N,1\right)  $ sigma model.

We see that, in addition to the possibility of a changing gravitational
constant after each phase transition, some general constraints have emerged
from 2T-Gravity on the structure of scalars in 1T field theory. The
constraints described in the above paragraphs, permeate to the shadow 1T field
theory in $d$ dimensions, and show up in the couplings among scalars in the
kinetic terms, potential energy $v\left(  s,n_{i}\right)  ,$ and gauge bosons
and fermions through the factors $\phi^{\frac{2\left(  d-4\right)  }{d-2}}$
and $\phi^{-\frac{d-4}{d-2}}$ respectively, thus leaving footprints that
observers in the conformal shadow in $d$ dimensions can use to infer
properties of the underlying 2T theory. These properties of scalars, including
the inflaton and the Higgs, are currently under investigation in cosmological
and LHC contexts \cite{2tcosmology}.

Additional and deeper observable properties of the 2T theory can be obtained
by studying the other shadows related to the conformal shadow by duality
transformations as in the examples in \cite{emergentfieldth1}%
\cite{emergentfieldth2} and their generalizations that are still under
investigation \cite{massiveGauge}.

\section{Kinematics of 2T curved space in $d+2$ dimensions}

\label{kinematicsd+2}

The equations of motion that follow from the 2T-gravity action in
Eq.(\ref{action}) and Table-1 can be divided into two categories: dynamical
equations and kinematical equations. The dynamical equations are those
proportional to the delta function $\delta\left(  W\right)  $ - these provide
the dynamics including the field interactions in $d+2$ dimensions. The
kinematical equations are those proportional to the derivatives of the delta
function $\delta^{\prime}\left(  W\right)  $ and $\delta^{\prime\prime}\left(
W\right)  $. We recall that such derivatives emerge from integration by parts
in computing the variation of the action.

A remarkable property of the kinematical equations that will be emphasized
below is that they are universal and have a geometrical character. They can be
shown to be independent of interactions and they are the same for each field
independent of which other fields are included in the action. Although these
properties may not be immediately apparent when the kinematic equations are
derived from the action, it follows after some rewriting of the equations as
seen below. There is an important underlying symmetry for this result, namely
Sp$\left(  2,R\right)  ,$ which will be discussed in section (\ref{eoms}). The
kinematical equations provide the instructions for how to relate the fields in
$\left(  d+2\right)  $ dimensions to the shadows in $d$ dimensions, so their
solutions reduce the original 2T theory to various shadows in $d$ dimensions,
such as the conformal shadow in Eq.(\ref{S0shad}).

Both the kinematic and dynamical equations for 2T-Gravity were derived from
the action in \cite{2Tgravity}. In this section we deal mainly with the
kinematics. For the pure gravity triplet $\left(  G_{MN},\Omega,W\right)  $
the kinematic equations have the following form
\begin{gather}
G^{MN}\partial_{M}W\partial_{N}W=4W,\;\;\label{kin1}\\
G^{MN}\partial_{M}W\partial_{N}\Omega=4a\Omega\left(  6-\nabla^{2}W\right)
,\label{kin2}\\
\nabla_{M}\partial_{N}W=G_{MN}\left[  -6+\nabla^{2}W+8a\left(  6-\nabla
^{2}W\right)  \right]  . \label{kin3}%
\end{gather}
where $\nabla$ is the covariant derivative in the curved space with metric
$G_{MN}.$ After contracting the third equation with $G^{MN}$ and taking
account of $\nabla^{2}W=G^{MN}\nabla_{M}\partial_{N}W,$ one can solve for
$\nabla^{2}W$ and find
\begin{equation}
\nabla^{2}W=\frac{6\left(  d+2\right)  \left(  8a-1\right)  }{\left(
d+2\right)  \left(  8a-1\right)  +1}=2\left(  d+2\right)  ,
\end{equation}
where the special value of $a$ in Eq.(\ref{a value}) is used. Note that the
result is independent of the metric, and in particular it is true in flat
space for $W_{flat}=X^{2}$ as listed in Table-1, namely $\left(  \nabla
^{2}W\right)  _{flat}=\eta^{MN}\partial_{M}\partial_{N}\left(  X^{2}\right)
=2\partial_{M}X^{M}=2\left(  d+2\right)  $.

With this result, the kinematic equations for the gravity triplet
(\ref{kin1}-\ref{kin3}) simplify to%
\begin{equation}
W=V\cdot V,\;\;V\cdot\partial\Omega=-\frac{d-2}{2}\Omega,\;\;G_{MN}=\nabla
_{M}V_{N}, \label{KinGrav}%
\end{equation}
where the dot products are constructed with $G^{MN}$ and the vector $V_{M}$ or
$V^{M}$ is defined as the derivative of $W\left(  X\right)  $
\begin{equation}
\;V_{M}\equiv\frac{1}{2}\partial_{M}W,\;V^{M}=\frac{1}{2}G^{MN}\partial_{N}W.
\label{V}%
\end{equation}
For this form of $V_{M}$ the expression for $G_{MN}=\frac{1}{2}\nabla
_{M}\partial_{N}W=\frac{1}{2}\partial_{M}\partial_{N}W-\frac{1}{2}\Gamma
_{MN}^{K}\partial_{K}W$ is symmetric $G_{MN}=\nabla_{M}V_{N}=\nabla_{N}V_{M}$
since the Christoffel symbol \cite{carrol}
\begin{equation}
\Gamma_{MN}^{K}=\frac{1}{2}G^{KQ}\left(  \partial_{M}G_{NQ}+\partial_{N}%
G_{MQ}-\partial_{Q}G_{MN}\right)  \label{christoffel}%
\end{equation}
is symmetric. In particular, in flat space all the kinematic equations above
are satisfied by
\begin{equation}
W_{flat}=X^{2},\;V_{M}^{flat}=X_{M},\;G_{MN}^{flat}=\eta_{MN},\;\left(
\Gamma_{MN}^{K}\right)  _{flat}=0 \label{flatW}%
\end{equation}
as listed in Table-1.

The form of the metric in (\ref{KinGrav}) that emerged from the 2T gravity
action satisfies a special geometric property. By using the definition of the
Lie derivative $\pounds _{V}$ for the vector $V,$ which on a tensor is given
by $\pounds _{V}G_{MN}\equiv\nabla_{M}V_{N}+\nabla_{N}V_{M},$ we can recognize
that the form $G_{MN}$ given in (\ref{KinGrav},\ref{V}) is equivalent to
writing the following \textit{homothety} equations for the metric and its
inverse%
\begin{equation}
\pounds _{V}G_{MN}=2G_{MN},\;\pounds _{V}G^{MN}=-2G^{MN}. \label{KinG}%
\end{equation}
The Lie derivative amounts to a general coordinate transformation of the
metric using the vector $V_{M}\left(  X\right)  $ as the parameter of
transformation, therefore we can say that under such a transformation the
metric yields a factor of $2$
\begin{align}
2G_{MN}  &  =\pounds _{V}G_{MN}=\nabla_{M}V_{N}+\nabla_{N}V_{M}\\
&  =V^{K}\partial_{K}G^{MN}+\partial_{M}V^{K}G_{KN}+\partial_{N}V^{K}%
G_{MK}.\nonumber
\end{align}
The equivalence of the homothety conditions above to the kinematic equations
of motion (\ref{KinGrav}) that emerged from the action is shown by inserting
the Christoffel symbol (\ref{christoffel}) into $\nabla_{M}V_{N}=\partial
_{M}V_{N}-\Gamma_{MN}^{K}V_{K}$.

After coupling the gravity triplet to any matter as in Table-1, the kinematic
equations initially derived from the action appear to have couplings between
the gravity triplet ($G_{MN},\Omega,W$) and matter fields \cite{2Tgravity}.
However, after using the kinematic equations for matter as well, one finds
that they simplify to the form above (\ref{KinGrav},\ref{V}) regardless of the
type of matter they couple to \cite{2Tgravity}. Furthermore, the kinematic
equations for matter fields $\left(  S_{i},\Psi,A_{M}\right)  $ also simplify
to the following form \cite{2Tgravity}%
\begin{equation}
V\cdot DS_{i}=-\frac{d-2}{2}S_{i},\;\;V\cdot D\Psi=-\frac{d}{2}\Psi
,\;\;V^{M}F_{MN}=0,\label{KinMatter}%
\end{equation}
where $F_{MN}=\partial_{M}A_{N}-\partial_{N}A_{M}-ig\left[  A_{M}%
,A_{N}\right]  $ is the Yang-Mills field strength, and $D_{M}$ is the
Yang-Mills covariant derivative. These equations can also be rewritten as the
response to the Lie derivative $\pounds _{V}$ applied on the corresponding
fields of various spins.

It is now evident that all kinematic equations (\ref{KinGrav},\ref{KinMatter})
derived from the action (\ref{action}) have a geometrical meaning and they are
the same for each field irrespective of the interactions and irrespective of
which other fields are included in the action. This is why we call these
\textquotedblleft kinematic\textquotedblright\ equations. The deeper
significance of this structure is an underlying Sp$\left(  2,R\right)  $
symmetry explained in section (\ref{eoms}).

\subsection{Kinematics of the metric, vielbein and Dirac gamma matrices}

The kinematic equations described above required the peculiar homothety
condition (\ref{KinG}) that the metric must satisfy $\pounds _{V}%
G_{MN}=2G_{MN}$, which in turn requires that the metric must also be
constructed from the potential $W\left(  X\right)  $
\begin{equation}
G_{MN}=\nabla_{M}V_{N},\;\text{with }V_{N}=\frac{1}{2}\partial_{N}W\text{ and
}W=V\cdot V \label{G2t}%
\end{equation}
This is a nonlinear equation since $\Gamma_{MN}^{P}$ is constructed from the
metric as in (\ref{christoffel}). These equations are solved by choosing
gauges and convenient coordinates. Then the solution is expressed in terms of
the shadow field $g_{\mu\nu}\left(  x\right)  $ in two lower dimensions, and
its prolongations, all of which remain arbitrary as far as the homothety
condition (\ref{G2t}) is concerned, as will be discussed below. In this
section we develop properties of the curved space described by a metric that
satisfies (\ref{G2t}) without choosing any gauges.

Before imposing the homothety condition (\ref{G2t}), we recall the well known
usual formulation of curved space, with any signature in any number of
dimensions. We define a base space and a tangent space. The vielbein that
connects the two spaces $E_{M}^{~a}\left(  X\right)  $ is labelled by an index
$M$ in base space and an index $a$ in tangent space. The metric in flat
tangent space is the SO$\left(  d,2\right)  $ invariant flat metric $\eta
_{ab}$ while the curved space metric $G_{MN}\left(  X\right)  $ is constructed
from the vielbein as a SO$\left(  d,2\right)  $ invariant%
\begin{equation}
G_{MN}\left(  X\right)  =E_{M}^{~a}\left(  X\right)  E_{N}^{~b}\left(
X\right)  \eta_{ab}. \label{metricE}%
\end{equation}
We introduce the usual affine connection $\Gamma_{MN}^{P}\left(  X\right)  $
of Eq.(\ref{christoffel}) which is symmetric in base space $\Gamma_{MN}%
^{P}=\Gamma_{NM}^{P}$, and the SO$\left(  d,2\right)  $ Yang-Mills field known
as the \textquotedblleft spin connection\textquotedblright\ $\omega_{M}%
^{~ab}\left(  X\right)  $ which is antisymmetric in tangent space $\omega
_{M}^{~ab}=-\omega_{M}^{~ba}$. We will use the following notation for various
covariant derivatives%
\begin{equation}
\nabla_{M}=\partial_{M}-\Gamma_{M}\;;\;D_{M}=\partial_{M}+\omega_{M}%
\;;\;\hat{D}_{M}=\partial_{M}-\Gamma_{M}+\omega_{M}\;. \label{der}%
\end{equation}
The first one $\nabla_{M}$ is covariant when applied on a field with only base
space indices, the second one $D_{M}$ is covariant when applied on a field
with only tangent space indices and the third one $\hat{D}_{M}$ is covariant
when applied on a field with both base and tangent space indices. In many
expressions we will write $\hat{D}_{M}$ and let it be understood that
sometimes $\Gamma_{M}$ or $\omega_{M}$ would drop out automatically depending
on the field. However, when it becomes useful we will specialize $\hat{D}_{M}$
to $\nabla_{M}$ or $D_{M}$ or even $\partial_{M}.$

Using these definitions, the covariant derivative of $E_{M}^{~a}$ that is
gauge invariant under general coordinate transformations as well as under the
tangent space local SO$\left(  d,2\right)  $ transformations is%
\begin{equation}
\hat{D}_{M}E_{N}^{~a}=\partial_{M}E_{N}^{~a}-\Gamma_{MN}^{~~~P}E_{P}%
^{~a}+\omega_{M}^{~ab}E_{Nb}^{~}\;. \label{covGW}%
\end{equation}
A symmetric connection $\Gamma_{MN}^{P}=\Gamma_{NM}^{P}$ demands vanishing
torsion $T_{MN}^{~a}$%
\begin{equation}
T_{MN}^{~a}\equiv\hat{D}_{M}E_{N}^{~a}-\hat{D}_{N}E_{M}^{~a}=\partial_{\lbrack
M}E_{N]}^{~a}+\omega_{\lbrack M}^{ab}E_{N]b}=D_{[M}E_{N]}^{~a}=0.
\label{torsion}%
\end{equation}
where $\Gamma_{MN}^{P}$ dropped out due to antisymmetrization. $T_{MN}^{~a}=0$
is an equation from which the spin connection $\omega_{M}^{~ab}$ is solved as
a function of $E_{M}^{a}$ as
\begin{equation}%
\begin{array}
[c]{c}%
\omega_{M}^{ab}=E^{Na}E^{Pb}\left(  C_{MNP}-C_{NPM}-C_{PMN}\right)  ,\\
C_{MNP}\equiv-\frac{1}{2}E_{Mc}\left(  \partial_{N}E_{P}^{c}-\partial_{P}%
E_{N}^{c}\right)  .
\end{array}
\label{spinConn}%
\end{equation}
Furthermore, the two connections $\Gamma_{MN}^{P}$ and $\omega_{M}^{~ab}$ are
related to each other by requiring that the covariant derivative of $E_{M}%
^{a}$ in Eq.(\ref{covGW}) vanishes%
\begin{equation}
\hat{D}_{M}E_{N}^{~a}=0\;\rightarrow\omega_{M}^{~ab}=\frac{1}{2}%
E^{~N[a}\left(  \nabla_{M}E_{N}^{b]}\right)  . \label{covEzero}%
\end{equation}
Eq.(\ref{covEzero}) insures that the covariant derivative of the metric
vanishes $\hat{D}_{P}G_{MN}=\nabla_{P}G_{MN}=0$, and since $\omega_{M}$ drops
out it can be written as%
\begin{equation}
\nabla_{P}G_{MN}=\partial_{P}G_{MN}-\Gamma_{PM}^{~~~Q}G_{QN}-\Gamma
_{PN}^{~~~Q}G_{MQ}=0.\; \label{Dgzero}%
\end{equation}
Then one can show that the $\Gamma_{MN}^{P}$ which solves both equations
(\ref{covEzero},\ref{Dgzero}) is nothing but the usual Christoffel connection
constructed from the metric $G_{MN}$ given in Eq.(\ref{christoffel}).

The curvature tensor is constructed just as in Yang-Mills theory from the spin
connection $\omega_{M}^{~ab}$ which is nothing but the Yang-Mills gauge field
for the SO$\left(  d,2\right)  $ local symmetry in tangent space
\begin{equation}
R_{MN}^{~ab}=\partial_{M}\omega_{N}^{~ab}-\partial_{N}\omega_{M}^{~ab}%
+\omega_{M}^{~ak}\omega_{Nk}^{~~~~b}-\omega_{N}^{~ak}\omega_{Mk}^{~~~~b}.
\label{YMR}%
\end{equation}
This Yang-Mills field strength coincides with the standard curvature tensor
$R_{QMN}^{P}$ constructed from the affine connection%
\begin{equation}
R_{QMN}^{P}=\partial_{M}\Gamma_{NQ}^{P}-\partial_{N}\Gamma_{MQ}^{M}%
+\Gamma_{MS}^{P}\Gamma_{NQ}^{S}-\Gamma_{NS}^{P}\Gamma_{MQ}^{S}%
\end{equation}
after converting the base indices to tangent indices%
\begin{equation}
R_{MN}^{~ab}=-R_{QMN}^{P}E_{P}^{a}E^{Qb}. \label{Rmnijpq}%
\end{equation}
As is well known, the curvature with all lower indices $G_{PK}R_{QMN}%
^{K}=R_{PQMN}$ is antisymmetric in $M\leftrightarrow N$ and separately under
$P\leftrightarrow Q,$ but is symmetric under the interchange of the pairs
$MN\leftrightarrow PQ,$ and satisfies the cyclic identity
\begin{equation}%
\begin{array}
[c]{c}%
R_{MNPQ}=-R_{NMPQ}=-R_{MNQP}=R_{PQMN},\;\\
R_{QMN}^{P}+R_{MNQ}^{P}+R_{NQM}^{P}=0.
\end{array}
\label{RmnpqSymm}%
\end{equation}

We now turn to the special kinematics of 2T-physics. The specialty in
2T-physics is that the metric is constructed from the covariant derivative of
the vector $V_{M}$ as in Eq.(\ref{G2t}). Applying the standard formalism
above, and imposing the homothety condition (\ref{G2t}), we obtain the
following seven lemmas that describe certain general properties of this
special gravitational system that are useful in our work:

\begin{enumerate}
\item The vielbein $E_{M}^{~a}$ is constructed from a vector $V^{a}$ in
tangent space
\begin{equation}
E_{M}^{~a}=D_{M}V^{a}=\partial_{M}V^{a}+\omega_{M}^{ab}V_{b}. \label{Emi}%
\end{equation}
where $W=V^{a}V_{a}$ and
\begin{equation}
V^{a}=V_{M}E^{Ma}=\frac{1}{2}\left(  \partial_{M}W\right)  E^{Ma},\;\text{or
}V_{M}=E_{M}^{a}V_{a}=\frac{1}{2}\left(  \partial_{M}W\right)  . \label{Vm}%
\end{equation}
This is shown by reconstructing the metric and using the following series of
steps to prove that it agrees with Eq.(\ref{G2t}) as follows
\begin{equation}
G_{MN}\left(  X\right)  =\left\{
\begin{array}
[c]{l}%
=E_{M}^{~a}\left(  X\right)  E_{N}^{~b}\left(  X\right)  \eta_{ab}=D_{M}%
V^{a}E_{N}^{~b}\left(  X\right)  \eta_{ab}\\
=\hat{D}_{M}\left(  V_{b}E_{N}^{~b}\left(  X\right)  \right)  =\hat{D}%
_{M}\left[  D_{N}\left(  \frac{1}{2}V^{a}V^{b}\eta_{ab}\right)  \right] \\
=\frac{1}{2}\nabla_{M}\partial_{N}W=\nabla_{M}V_{N}.
\end{array}
\right\}
\end{equation}
In going from the first to the second line we used $\hat{D}_{M}E_{N}^{~b}=0$
of Eq.(\ref{covEzero}), the rest of the steps are evident. Hence the structure
of the vielbein in (\ref{Emi}) is equivalent to the homothety condition
(\ref{G2t}) on $G_{MN}.$

\item The vanishing of torsion $T_{MN}^{a}=0$ requires the following kinematic
conditions on the curvature
\begin{equation}
R_{MN}^{~ab}V_{b}=0,\;\;V^{P}R_{PQMN}=0,\;\;R_{MNPQ}V^{P}=0,\;V^{M}%
R_{MN}^{~ab}=0. \label{L1}%
\end{equation}
The first form is shown by inserting $E_{N}^{~a}=\hat{D}_{N}V^{a}$ in the
vanishing torsion $0=T_{MN}^{~a}\equiv\hat{D}_{[M}E_{N]}^{~a}=\left[  \hat
{D}_{M},\hat{D}_{N}\right]  V^{a}=R_{MN}^{ab}V_{b}.$ The second form follows
from the first by replacing tangent indices by base indices, or directly by
writing the vanishing torsion in the form $T_{MN}^{P}=\left[  \nabla
_{M},\nabla_{N}\right]  V^{P}=V^{Q}R_{QMN}^{P}=0.$ The third form follows from
the second by using the identity $R_{PQMN}=R_{MNPQ}.$ The last form follows
from converting the $P,Q$ indices to $ab$ indices in the third form. It should
be noted that the last form $V^{M}R_{MN}^{~ab}=0$ is the standard kinematic
equation required by Sp$\left(  2,R\right)  $ constraints on any Yang-Mills
field strength for any gauge group $V^{M}F_{MN}^{~a}$ as given in
Eq.(\ref{KinMatter}).

\item The SO$\left(  d,2\right)  $ Dirac gamma matrices with base space
indices $\gamma_{M}\equiv E_{M}^{a}\gamma_{a}$ are covariantly constant%
\begin{equation}
\hat{D}_{M}\gamma_{N}=\partial_{M}\gamma_{N}-\Gamma_{MN}^{P}\gamma_{P}%
+\frac{1}{4}\omega_{M}^{ab}\left(  \gamma_{ab}\gamma_{N}-\gamma_{N}\gamma
_{ab}\right)  =0\; \label{L2}%
\end{equation}
Here the covariant derivative $\hat{D}_{M}$ includes $\omega_{M}^{ab}$ because
$\left(  \gamma_{M}\right)  _{A\dot{B}}$ has spinor indices $A\dot{B}$ in
tangent space. To show this result, consider first the tangent space gamma
matrices $\gamma_{a},$ which are pure constants that satisfy $\partial
_{M}\gamma_{a}=0$. For these the covariant derivative also gives $\hat{D}%
_{M}\gamma_{a}=0$ because it reduces to the ordinary derivative
\begin{equation}
\hat{D}_{M}\gamma_{a}=D_{M}\gamma_{a}=\partial_{M}\gamma_{a}=0,
\end{equation}
because the $\omega_{M}^{ab}$ contributions for all tangent space indices
$a,A,\dot{B}$ in $\left(  \gamma_{a}\right)  _{A\dot{B}}$ cancel each other.
Then the result in Eq.(\ref{L2}) is shown by rewriting $\hat{D}_{M}\gamma
_{N}=\hat{D}_{M}\left(  E_{N}^{a}\gamma_{a}\right)  =\left(  \hat{D}_{M}%
E_{N}^{a}\right)  \gamma_{a}=0$ which follows from (\ref{covEzero}).

\item The covariant derivative of the gamma matrix $V=V^{N}\gamma_{N}%
=V^{a}\gamma_{a},$ which appears in the Yukawa couplings in Table-1, gives
$\gamma_{M}$%
\begin{equation}
\hat{D}_{M}V=D_{M}V=\gamma_{M}. \label{L3}%
\end{equation}
This is shown by writing $D_{M}V=\left(  D_{M}V^{a}\right)  \gamma_{a}%
=E_{M}^{a}\gamma_{a}=\gamma_{M}.$

\item The ordinary derivative of $V^{2}=V^{a}V_{a}=V^{M}V_{M}$ gives $2V_{M}$%
\begin{equation}
\partial_{M}V^{2}=2V_{M} \label{L4}%
\end{equation}
This is shown by writing $\partial_{M}V^{2}=D_{M}V^{2}=2\left(  D_{M}%
V^{a}\right)  V_{a}=2E_{M}^{~a}V_{a}=2V_{M}.$ Of course, this is in agreement
with the fact that $W=V^{2}$ and the definition $V_{M}=\frac{1}{2}\partial
_{M}W.$

\item The various fields $V_{a},V_{M},V$ automatically satisfy the following
kinematic equations
\begin{equation}
\left(  V^{M}D_{M}-1\right)  V_{a}=0,\;\left(  V^{M}\nabla_{M}-1\right)
V_{N}=0,\;\left(  V^{M}D_{M}-1\right)  V=0. \label{L5}%
\end{equation}
These follow from $D_{M}V_{a}=E_{Ma},\ \nabla_{M}V_{N}=G_{MN}$ and
$D_{M}V=\gamma_{M}$ derived above.

\item The following kinematic property in $d+2$ dimensions is automatically
satisfied
\begin{equation}
D_{M}\left(  \sqrt{G}\delta\left(  V^{2}\right)  \gamma^{M}V\right)
=d\sqrt{G}\delta\left(  V^{2}\right)  \label{L7}%
\end{equation}
To show this, first recall that the divergence of any vector $\nabla_{M}%
v^{M}=\partial_{M}v^{M}+\Gamma_{MP}^{~~~M}v^{P}$ can be rewritten as
$\nabla_{M}v^{M}=G^{-1/2}\partial_{M}(\sqrt{G}v^{M}).$ Applying this to the
vector $v^{M}\equiv\delta\left(  V^{2}\right)  \gamma^{M}V$, gives
$D_{M}(\sqrt{G}v^{M})=\sqrt{G}\hat{D}_{M}v^{M}$ where $\hat{D}_{M}$ appears.
Then use the properties derived in the lemmas above as follows%
\begin{align*}
&  D_{M}\left[  \sqrt{G}\delta\left(  V^{2}\right)  \gamma^{M}V\right] \\
&  =\sqrt{G}\hat{D}_{M}\left[  \delta\left(  V^{2}\right)  \gamma^{M}V\right]
\\
&  =\sqrt{G}\left(  \hat{D}_{M}\delta\left(  V^{2}\right)  \right)  \gamma
^{M}V+\sqrt{G}\delta\left(  V^{2}\right)  \left(  \hat{D}_{M}\gamma
^{M}\right)  V+\sqrt{G}\delta\left(  V^{2}\right)  \gamma^{M}\hat{D}_{M}V\\
&  =\sqrt{G}\delta^{\prime}\left(  V^{2}\right)  \left(  2V_{M}\right)
\gamma^{M}V+\sqrt{G}\delta\left(  V^{2}\right)  \left(  0\right)  V+\sqrt
{G}\delta\left(  V^{2}\right)  \gamma^{M}\gamma_{M}\\
&  =\sqrt{G}\delta^{\prime}\left(  V^{2}\right)  2V^{2}+\sqrt{G}\delta\left(
V^{2}\right)  \left(  d+2\right) \\
&  =d\sqrt{G}\delta\left(  V^{2}\right)
\end{align*}
To get to the last step we have used the property of the delta function
$V^{2}\delta^{\prime}\left(  V^{2}\right)  =-\delta\left(  V^{2}\right)  .$
\end{enumerate}

\section{Dynamical and kinematic equations of motion}

\label{eoms}

The dynamical equations of motion derived from the action (\ref{action}), and
its generalization from Table-1, are those proportional to $\delta\left(
W\right)  $ for the general variation of every field. The dynamical equations
that follow from varying the metric, dilaton and scalars are \cite{2Tgravity}
(neglecting fermions and gauge bosons)%
\begin{align}
\delta\Omega &  :\;\left[  \nabla^{2}\Omega-2aR\Omega+a\partial_{\Omega
}V\left(  \Omega,S_{i}\right)  \right]  _{W=0}=0\label{d1}\\
\delta S_{i} &  :\;\left[  \nabla^{2}S_{i}-2aRS_{i}-\partial_{S_{i}}V\left(
\Omega,S_{i}\right)  \right]  _{W=0}=0\label{d2}\\
\delta G^{MN} &  :\;\left[  R_{MN}\left(  G\right)  -\frac{1}{2}G_{MN}R\left(
G\right)  -T_{MN}\right]  _{W=0}=0\label{d3}%
\end{align}
where the stress tensor $T_{MN}$ is%
\begin{equation}
T_{MN}=\frac{1}{\Omega^{2}-aS_{i}^{2}}\left[
\begin{array}
[c]{l}%
\frac{-1}{2a}\partial_{M}\Omega\partial_{N}\Omega+\frac{1}{2}\partial_{M}%
S_{i}\partial_{N}S_{i}\\
+G_{MN}\left(  \frac{1}{4a}\left(  \partial\Omega\right)  ^{2}-\frac{1}%
{4}\left(  \partial S_{i}\right)  ^{2}-\frac{1}{2}V\left(  \Omega
,S_{i}\right)  \right)  \\
-\left(  G_{MN}\nabla^{2}-\nabla_{M}\partial_{N}\right)  \left(  \Omega
^{2}-aS_{i}^{2}\right)
\end{array}
\right]  \label{stress}%
\end{equation}
and as usual $R_{MN}\left(  G\right)  \equiv R_{MPN}^{P}$ and $R\left(
G\right)  \equiv G^{MN}R_{MN}.$ These equations are to be solved at $W=0$
because of the delta function $\delta\left(  W\right)  $ that multiplies them,
but we will at first manipulate them for any $W$.

We now simplify these equations as follows. Contracting Eq.$\left(
\text{\ref{d3}}\right)  $ with $G^{MN},$ we can solve for $R\left(  G\right)
=\frac{-2}{d}G^{MN}T_{MN}$ and get%
\begin{equation}
\left(  \Omega^{2}-aS_{i}^{2}\right)  R\left(  G\right)  =\left[
\begin{array}
[c]{l}%
-\frac{1}{2a}\partial\Omega\cdot\partial\Omega+\frac{1}{2}\partial S_{i}%
\cdot\partial S_{i}\\
+\frac{2\left(  d+1\right)  }{d}\nabla^{2}\left(  \Omega^{2}-aS_{i}%
^{2}\right)  +\frac{d+2}{d}V\left(  \Omega,S_{i}\right)
\end{array}
\right]  \label{R0}%
\end{equation}
Multiply Eqs.(\ref{d1},\ref{d2}) by $\left(  -\Omega/a\right)  ,S_{i}$
respectively, sum over $i$ and add them, to get
\begin{equation}
0=2R\left(  \Omega^{2}-aS_{i}^{2}\right)  -\frac{1}{a}\left(  \Omega\nabla
^{2}\Omega-aS_{i}\nabla^{2}S_{i}\right)  -\left(  \Omega\partial_{\Omega
}+S_{i}\partial_{S_{i}}\right)  V\left(  \Omega,S\right)  .
\end{equation}
In this equation we insert the expression in (\ref{R0}) and use the
homogeneity of the potential (\ref{Vt}) to write $\left(  \Omega
\partial_{\Omega}+S_{i}\partial_{S_{i}}\right)  V\left(  \Omega,S\right)
=\frac{2d}{d-2}V\left(  \Omega,S\right)  ,$ and after some simplifications we
obtain
\begin{equation}
\nabla^{2}\left(  \Omega^{2}-aS_{i}^{2}\right)  =-V\left(  \Omega
,S_{i}\right)  . \label{d2WV}%
\end{equation}
Inserting this back into (\ref{R0}) yields
\begin{equation}
R\left(  G\right)  =\frac{-\frac{1}{2a}\partial\Omega\cdot\partial\Omega
+\frac{1}{2}\partial S_{i}\cdot\partial S_{i}-V\left(  \Omega,S_{i}\right)
}{\Omega^{2}-aS_{i}^{2}}. \label{R}%
\end{equation}
Using both Eqs.(\ref{d2WV},\ref{R}), the energy momentum tensor in
Eq.(\ref{stress}) simplifies to%
\begin{equation}
T_{MN}=\frac{1}{\left(  \Omega^{2}-aS_{i}^{2}\right)  }\left[
\begin{array}
[c]{c}%
-\frac{1}{2a}\partial_{M}\Omega\partial_{N}\Omega+\frac{1}{2}\partial_{M}%
S_{i}\partial_{N}S_{i}\\
-\frac{1}{2}G_{MN}\left(  \Omega^{2}-aS_{i}^{2}\right)  R+\nabla_{M}%
\partial_{N}\left(  \Omega^{2}-aS_{i}^{2}\right)
\end{array}
\right]  \label{TMN}%
\end{equation}
Inserting (\ref{R},\ref{TMN}) into (\ref{d3}) yields
\begin{equation}
R_{MN}\left(  G\right)  =S_{MN}\left(  \Omega,S_{i}\right)  , \label{RS}%
\end{equation}
where $S_{MN}\left(  \Omega,S_{i}\right)  $ is given by%
\begin{equation}
S_{MN}\left(  \Omega,S_{i}\right)  \equiv\frac{1}{\left(  \Omega^{2}%
-aS_{i}^{2}\right)  }\left[  -\frac{1}{2a}\partial_{M}\Omega\partial_{N}%
\Omega+\frac{1}{2}\partial_{M}S_{i}\partial_{N}S_{i}+\nabla_{M}\partial
_{N}\left(  \Omega^{2}-aS_{i}^{2}\right)  \right]  . \label{SMN}%
\end{equation}
Of course, $T_{MN}$ and $S_{MN}$ are related by $T_{MN}=S_{MN}-\frac{1}%
{2}G_{MN}G^{PQ}S_{PQ}.$ This is as much as we can simplify the dynamical
equations before choosing gauges and imposing $W=0.$

We also gather the kinematic equations satisfied by these fields and $W$ as
discussed in the previous section, with $V_{M}\equiv\frac{1}{2}\partial
_{M}W.$
\begin{gather}
W=V\cdot V,\;G_{MN}=\nabla_{M}V_{N},\;V^{P}R_{PQMN}=0,\;V^{M}R_{MN}%
=0,\label{k1}\\
\;\;V\cdot\partial\Omega=-\frac{d-2}{2}\Omega,\;\;V\cdot\partial S_{i}%
=-\frac{d-2}{2}S_{i},\;V^{M}S_{MN}=0.\label{k2}%
\end{gather}
A remarkable property is that the variation of the action with respect to $W$
does not give a new equation besides those kinematic or dynamical equations
that are obtained from the variation of the other fields. This was explained
\cite{2Tgravity} as being due to a local symmetry that allows $W\left(
X\right)  $ to be set to any desired function of $X^{M}.$ Although $W$ is set
to zero eventually in the dynamical equations (\ref{d1}-\ref{d3}), its first
and second derivatives that are related to $V_{M}$ and $G_{MN}$ do not vanish
(see e.g. the flat case in Eq.(\ref{flatW})). Exercising the freedom in
choosing some $W\left(  X\right)  $ is one of the steps that defines the
shadow in lower dimensions. The selection that leads to the conformal shadow
will be described in the next section.

\subsection{The underlying Sp$\left(  2,R\right)  $}

In the previous section we showed that the kinematic equations have a
geometrical significance. Now we emphasize that both the kinematic and
dynamical equations are intimately related to the fundamental Sp$\left(
2,R\right)  $ gauge symmetry that is at the root of 2T-physics. The
significance of the kinematic equations is that they impose part of the
\textit{gauge invariant physical state} conditions under Sp$\left(
2,R\right)  $ which is explained as follows. It was shown in \cite{2Tgravity}
that the three generators $Q_{ij}$ of Sp$\left(  2,R\right)  $ in the presence
of gravity are given by the following three functions of phase space $\left(
X^{M},P_{M}\right)  $
\begin{equation}
Q_{11}=W\left(  X\right)  ,\;Q_{12}=Q_{21}=V^{M}\left(  X\right)
P_{M},\;Q_{22}=G^{MN}\left(  X\right)  P_{M}P_{N}.\label{sp2rQ}%
\end{equation}
These $Q_{ij}$ form the Sp$\left(  2,R\right)  $ Lie algebra under Poisson
brackets provided the fields $W\left(  X\right)  ,V^{M}\left(  X\right)
,G^{MN}\left(  X\right)  $ satisfy the kinematic equations in
Eqs.(\ref{KinGrav},\ref{V},\ref{KinG}). The reader can check that in flat
space $W_{flat}=X^{2},$ $V_{flat}^{M}=X_{M}$ and $G_{flat}^{MN}=\eta^{MN}$
satisfy the Sp$\left(  2,R\right)  $ closure property under Poisson brackets.
These $Q_{ij}$ generate a \textit{local gauge symmetry on the worldline} for a
particle interacting with gravity, thus making its position and momentum
$X^{M}\left(  \tau\right)  ,P_{M}\left(  \tau\right)  $ indistinguishable at
every worldline instance \cite{2Tgravity}. In the quantum theory of such a
particle, its physical states must be Sp$\left(  2,R\right)  $ gauge
invariant, and hence these $Q_{ij}$ must vanish on the first quantized
wavefunctions. In position space the first quantized wavefunctions are the
fields in 2T field theory. Therefore these fields must satisfy $Q_{ij}\sim0$
after a proper quantum ordering of $X,P,$ and replacing the momentum by a
derivative $P_{M}=-i\partial_{M}.$ The kinematic equations in (\ref{KinGrav}%
,\ref{V},\ref{KinG}) imposed by the action are the precise expressions of the
vanishing of the generator $Q_{12}=\left(  -iV^{M}\partial_{M}+\cdots\right)
\sim0$ after appropriate quantum ordering for matter or gravitational fields
of various spins. The vanishing of $Q_{11}=W\left(  X\right)  $ is imposed
through the delta function $\delta\left(  W\right)  $ and its derivatives, and
finally the vanishing of $Q_{22}=\left(  -G^{MN}\partial_{M}\partial
_{N}+\cdots\right)  $ amounts to the \textit{dynamical} equations of
motion\footnote{The dots \textquotedblleft$\cdots$\textquotedblright\ in the
expressions of $Q_{ij}$ are the corrections due to interactions. This general
property is explained in refs.\cite{2tbrst2006}\cite{2tstandardM}%
\cite{2Tgravity}. These corrections, in the case of gravity, are precisely
supplied directly by the action in Eq.\ref{action} and Table-1, so they are
determined and written out fully in the kinematic (e.g. \ref{KinMatter}) and
dynamical equations (e.g. \ref{d1}-\ref{d3}) discused in this paper as well as
ref.\cite{2Tgravity}.}. Thus we see that all the equations of motion generated
by the 2T-field theory have the significance of imposing the physical state
condition under the Sp$\left(  2,R\right)  $ gauge symmetry, or more
precisely, its extension that includes particles with spin as well as
interactions, as explained in \cite{2tbrst2006}\cite{2tstandardM}.

One additional point of clarification about the role of the underlying
Sp$\left(  2,R\right)  ,$ as reflected in the kinematics, is in order. The
BRST field theory formulation in \cite{2tbrst2006} is technically a fuller
approach for imposing Sp$\left(  2,R\right)  ,$ but the extra baggage of the
BRST formalism, in the form of ghosts and redundant gauge degrees of freedom,
can be avoided by appreciating a few simple aspects related to Sp$\left(
2,R\right)  $ as just outlined in the previous paragraph. A related point is
that the underlying Sp$\left(  2,R\right)  $ provides the key for the
resolution of an ambiguity about the kinematic equations as derived from the
action (\ref{action}) and Table-1. This ambiguity is avoided through the BRST
approach, but is more easily resolved directly as follows. The variation of
the action for each field yields a linear superposition of the delta function
and its derivatives of the form $A\delta\left(  W\right)  +B\delta^{\prime
}\left(  W\right)  +C\delta^{\prime\prime}\left(  W\right)  =0.$ These imply
three equations that are satisfied at $W=0,$ but there is ambiguity in
identifying the proper forms of $A,B,C$ that should vanish at $W=0$. This is
because these distributions satisfy $W\delta^{\prime\prime}\left(  W\right)
=-2\delta^{\prime}\left(  W\right)  $ and $W\delta^{\prime}\left(  W\right)
=-\delta\left(  W\right)  .$ Therefore, if we add to $B$ a term that is
proportional to $W,$ that term feeds into a term added to $A.$ Similarly any
terms proportional to $W$ and $W^{2}$ in $C$ feed into $B$ and $A$
respectively. In the BRST approach the ambiguities of adding such terms to $B$
or $C$ are just gauge degrees of freedom which in any case drop out
automatically in the physical sector. When the BRST approach is
short-circuited as explained in \cite{2tstandardM}, this ambiguity is resolved
by recognizing that the $B=C=0$ kinematic equations amount to demanding the
closure of the underlying Sp$\left(  2,R\right)  $ Lie algebra, as made clear
by the $Q_{ij}$ in Eq.(\ref{sp2rQ}) for the corresponding worldline particle
model \cite{2Tgravity}. This closure demands that the equations $B=C=0$ must
be valid for all $W,$ not only $W=0,$ so that Sp$\left(  2,R\right)  $ is
defined and its Lie algebra is satisfied without restrictions on the phase
space degrees of freedom. This is necessary for it to be a gauge symmetry of
the particle model. The upshot is that the particle model can be used as a
guide to identify the correct forms of $B,C$ and then demand $B=C=0$ not only
at $W=0$ but at all $W,$ which means that if $B,C$ are expanded in powers of
$W,$ the coefficient of each power of $W$ should vanish. This is a shortcut to
insure self consistency of all the equations of motion, including the
dynamical equations, derived from the action (i.e. consistency of having first
class constraints $Q_{11},Q_{12},Q_{22}$, which then are set to zero). By
satisfying Sp$\left(  2,R\right)  $ in this way, the ambiguities in $A,B,C$
are resolved at any $W$. This insures the validity of the underlying
Sp$\left(  2,R\right)  $ gauge symmetry, and turn the ambiguities into gauge
freedom, consistent with the BRST approach \cite{2tstandardM}. Thus, the
physical sector that is gauge invariant under Sp$\left(  2,R\right)  ,$ namely
$B=C=0$ at any $W,$ and $A|_{W=0}=0,$ are the consistent field equations of motion.

Accordingly, it should be emphasized that the kinematic equations above
(\ref{k1},\ref{k2}), which are consistent with the particle model
\cite{2Tgravity}, are to be solved at any $W,$ not only at $W=0,$ while the
dynamical equations (\ref{d1}-\ref{SMN}) need to be satisfied only at $W=0.$
This is the procedure followed in the following sections to obtain the
conformal shadow and its prolongation.

The same result is also obtained without using the guidance of the particle
model discussed in the two previous paragraphs, but only using the gauge
symmetry in the equations of motion $A\delta\left(  W\right)  +B\delta
^{\prime}\left(  W\right)  +C\delta^{\prime\prime}\left(  W\right)  =0$ that
follow from the 2T field theory action. To explain this gauge symmetry we will
make a coordinate transformation, $X^{M}\rightarrow\left(  w,u,x^{\mu}\right)
,$ such that $W\left(  X\right)  =w$ is one of the coordinates, as in the next
section. Furthermore, to simplify the discussion we will concentrate only on a
single scalar field, say the dilaton $\Omega\left(  w,u,x^{\mu}\right)  ,$ and
suppress the coordinates $u,x^{\mu}$ since they are irrelevant to the
discussion. A similar discussion will hold for each field in the theory.

We want to show that the action has a gauge symmetry under the gauge
transformation $\delta_{\Lambda}\Omega=\Lambda_{\Omega}\left(  w,u,x\right)  $
for off shell arbitrary $\Omega$ as well as off shell other fields. The
variation of the action with respect to the field $\Omega$ takes the form%
\begin{equation}
\delta_{\Lambda}S=\int dwdud^{d}x\delta_{\Lambda}\Omega\left(  w\right)
\left[  A_{\Omega}\left(  w\right)  \delta\left(  w\right)  +B_{\Omega}\left(
w\right)  \delta^{\prime}\left(  w\right)  +C_{\Omega}\left(  w\right)
\delta^{\prime\prime}\left(  w\right)  \right]
\end{equation}
Of course, $A_{\Omega},B_{\Omega},C_{\Omega}$ depend on $w$ through $\Omega$
and other fields as well. Due to the delta functions we need to analyze the
expansion of each term in powers of $w$ and then do the integral over $w.$
Hence we have%
\begin{align}
\Omega\left(  w\right)   &  =\Omega_{0}+w\Omega_{1}+\frac{1}{2}w^{2}\Omega
_{2}+\cdots\label{expandO}\\
A_{\Omega}\left(  w\right)   &  =A_{0}+wA_{1}+\frac{1}{2}w^{2}A_{2}+\cdots\\
B_{\Omega}\left(  w\right)   &  =B_{0}+wB_{1}+\frac{1}{2}w^{2}B_{2}+\cdots\\
C_{\Omega}\left(  w\right)   &  =C_{0}+wC_{1}+\frac{1}{2}w^{2}C_{2}+\cdots\\
\Lambda_{\Omega}\left(  w\right)   &  =\Lambda_{0}+w\Lambda_{1}+\frac{1}%
{2}w^{2}\Lambda_{2}+\cdots
\end{align}
Then the integral gives%
\begin{equation}
\delta_{\Lambda}S=\int dud^{d}x\left[  \Lambda_{0}\left(  A_{0}-B_{1}%
+C_{2}\right)  +\Lambda_{1}\left(  -B_{0}+2C_{1}\right)  +\Lambda_{2}%
C_{0}\right]  .
\end{equation}
It is possible to make $\delta_{\Lambda}S=0$ with a choice of gauge parameters
$\Lambda_{0},\Lambda_{1},\Lambda_{2}$ that are related to each other, when all
the fields are off-shell. There are three local parameters but only one
condition, hence two of the parameters among $\Lambda_{0},\Lambda_{1}%
,\Lambda_{2}$ can be chosen arbitrarily such that the action is gauge
invariant $\delta_{\Lambda}S=0$ off-shell. This 2-parameter gauge symmetry is
a remnant of the Sp$\left(  2,R\right)  $ BRST gauge symmetry discussed in
\cite{2tbrst2006}. A similar local symmetry is valid separately for each field
in any 2T-field theory. This was called the 2T-gauge symmetry in
\cite{2tstandardM}.

Using this gauge symmetry we can choose arbitrarily the prolongations
$\Omega_{1}\left(  u,x\right)  $ and $\Omega_{2}\left(  u,x\right)  $ in the
expansion of Eq.(\ref{expandO}). It is convenient to make the choice of
$\Omega_{1},\Omega_{2}$ such that $B_{1}=C_{2}$ and $C_{1}=0.$ These gauge
choices hold when all the fields are off-shell.

Now we investigate the on-shell equations of motion which are obtained from
the above procedure by taking $\delta\Omega_{0},\delta\Omega_{1},\delta
\Omega_{2}$ arbitrary and independent of each other. So the equations of
motion for the on-shell $\Omega_{0,1,2}$ are%
\begin{equation}
A_{0}-B_{1}+C_{2}=0,\;-B_{0}+2C_{1}=0,\;C_{0}=0.
\end{equation}
In the gauge we have chosen they become $A_{0}=0,\;B_{0}=0,\;B_{1}%
=C_{2},\;C_{0}=0,\;C_{1}=0.$

Now we investigate what $C_{2}$ is in more detail. It was shown in
\cite{2Tgravity} that in the variation of the action with respect to every
field the term $C\delta^{\prime\prime}\left(  W\right)  $ is always of the
form $C\sim\left(  G^{MN}\partial_{M}W\partial_{N}W-4W\right)  $ up to a field
dependent proportionality factor. In the next section we show that in the
coordinate system $W\left(  X\right)  =w,$ this expression becomes zero
automatically by constraining only the $G^{ww}$ component of the metric
$G^{MN}\left(  X\right)  .$ Therefore, we automatically obtain $C_{2}=0.$

With this result for $C_{2}=0$ taken into account, we now see that, in our
chosen gauge, the on-shell dynamics must satisfy
\begin{equation}
A_{0}=0,\;B_{0}=0,\;B_{1}=0,\;C_{0}=0,\;C_{1}=0,\;C_{2}=0.
\end{equation}
The coefficients of the higher powers of $w$ in the expansion of $A_{\Omega
}\left(  w\right)  ,B_{\Omega}\left(  w\right)  ,C_{\Omega}\left(  w\right)
,$ such as $A_{n\geq1},$ $B_{n\geq2}$ and $C_{n\geq3}$ are arbitrary because
they never enter in the equations. So they could be chosen arbitrarily without
any consequence for the dynamics of the fields $\Omega_{0},\Omega_{1}%
,\Omega_{2}$ which do appear in the equations. In particular, imposing
$B_{n\geq2}=0$ and $C_{n\geq3}=0$ has no consequences for the field components
$\Omega_{0},\Omega_{1},\Omega_{2}$ since they only restrict $\Omega_{n\geq3}.$
The latter are pure gauge freedom which never appear in the equations or even
in the off-shell action. Similar statements apply to the other fields.

This is in agreement with the procedure we discussed above, of solving the
equations $A\delta\left(  W\right)  +B\delta^{\prime}\left(  W\right)
+C\delta^{\prime\prime}\left(  W\right)  =0$ for all the fields by imposing
$A|_{W=0}$ while taking $B=C=0$ at all $W$ consistently with Sp$\left(
2,R\right)  .$ As we have shown, this is the consequence of a gauge choice,
consistent with the gauge symmetries of the action in Eq.(\ref{action}), as
well as with the Sp$\left(  2,R\right)  $ gauge symmetry properties of the
worldline formulation of particle dynamics in the presence of gravity.

\section{General Relativity as a shadow with Weyl symmetry}

\label{shadow}

In this section we determine the shadow fields and their prolongations. For
scalar fields $\Omega,S_{i}\left(  X\right)  ,$ these are defined by expanding
the field in powers of $W\left(  X\right)  ,$ as done below. The zeroth order
term is the shadow. The coefficients of all higher powers are Kaluza-Klein
type degrees of freedom, which we call prolongations of the shadow. For fields
that have spin indices, such as $G_{MN}\left(  X\right)  ,R_{MNPQ}\left(
X\right)  ,$ the zeroth order term has components that point in two lower
dimensions, such as $g_{\mu\nu},$ $R_{\mu\nu\lambda\sigma},$ as well as
components that point in the additional two dimensions. In traditional
Kaluza-Klein terminology the extra components are additional KK degrees of
freedom. In our case all such KK-type degrees of freedom, as well as the
coefficients of the higher powers in $W$ are called prolongations.

We will take advantage of gauge symmetries to eliminate some of the redundant
gauge degrees of freedom to clearly identify the physical degrees of freedom
recognized in 1T field theory in $d$ dimensions. The result will be that the
fields in $d+2$ dimensions $G_{MN}\left(  X\right)  ,\Omega\left(  X\right)
,S_{i}\left(  X\right)  $ will be reduced to the fields in $d$ dimensions
$g_{\mu\nu}\left(  x\right)  ,\phi\left(  x\right)  ,s_{i}\left(  x\right)  $
by a series of steps that involve gauge fixing as well as solving the
kinematic equations. The prolongations of the shadows $g_{\mu\nu}\left(
x\right)  ,\phi\left(  x\right)  ,s_{i}\left(  x\right)  ,$ from the
\textquotedblleft wall\textquotedblright\ $x^{\mu}$ into the higher
dimensional space $X^{M},$ namely the full $G_{MN}\left(  X\right)
,\Omega\left(  X\right)  ,S_{i}\left(  X\right)  ,$ will also be discussed. In
this section the shadows and their prolongations will be allowed to be
arbitrary fields in $d$ dimensions, restricted only by the kinematic
conditions, but in the following section, by using the dynamical equations of
the full theory it will be shown that all prolongations become functions of
only the shadow fields $g_{\mu\nu}\left(  x\right)  ,\phi\left(  x\right)
,s_{i}\left(  x\right)  $. One of the goals in this section is to show that
the accidental Weyl symmetry of Eq.(\ref{rescale}) acting on $g_{\mu\nu
}\left(  x\right)  ,\phi\left(  x\right)  ,s_{i}\left(  x\right)  $ in General
Relativity in the shadow action (\ref{S0shad2}) emerges from the local
coordinate reparametrization symmetry in the higher spacetime $X^{M}.$ It will
be clarified how a generalization of the Weyl symmetry acts also on the prolongations.

Among the local symmetries in 2T-gravity there are obviously general
coordinate transformations and the local symmetry that allows arbitrary
transformation of $W$ \cite{2Tgravity} as emphasized above. Exercising the
freedom of making gauge choices for these local symmetries defines the
properties of the emergent spacetimes for the shadows in the lower dimensions.

To begin this process we parametrize the spacetime $X^{M}$ in terms of $d+2$
coordinates $\left(  w,u,x^{\mu}\right)  $ and define the tangent basis in
base space $\partial_{M}=\left(  \partial_{w},\partial_{u},\partial_{\mu
}\right)  $ relative to these coordinates. In this basis we use the general
coordinate transformations to gauge fix $d+2$ components of the metric,
$G^{w\nu}=0,G^{wu}=-1,G^{uu}=0,$ leading to the following gauge fixed form of
$G^{MN}(X)$%
\begin{equation}
G^{MN}=%
\begin{array}
[c]{cc}%
M\backslash N &
\begin{array}
[c]{ccc}%
w & \text{ \ \ }u\text{ \ } & \text{ \ }\nu
\end{array}
\\%
\begin{array}
[c]{c}%
w\\
u\\
\mu
\end{array}
& \left(
\begin{array}
[c]{ccc}%
G^{ww} & -1 & 0\\
-1 & 0 & G^{u\nu}\\
0 & G^{\mu u} & G^{\mu\nu}%
\end{array}
\right)
\end{array}
\end{equation}
Next we select $W\left(  w,u,x^{\mu}\right)  =w$ to be simply one of the
coordinates, which immediately gives $V_{M}=\frac{1}{2}\partial_{M}W=\left(
\frac{1}{2},0,0\right)  _{M}.\;$Inserting this in the kinematic equation
$W=G^{MN}V_{M}V_{N}$ gives $w=\frac{1}{4}G^{ww}$ which fixes another component
of the metric. The result of these steps is then%
\begin{equation}
W\left(  X\right)  =w,\;\;G^{ww}\left(  X\right)  =4w,\;\;V_{M}\left(
X\right)  =\left(  \frac{1}{2},0,0\right)  _{M},\;\;V^{M}\left(  X\right)
=\left(  2w,-\frac{1}{2},0\right)  ^{M}. \label{gaugeW}%
\end{equation}

This choice of $W$ gives $V^{M}\partial_{M}=2w\partial_{w}-\frac{1}{2}%
\partial_{u},$ and the kinematic conditions for the scalars $\Omega,S_{i}$ in
(\ref{k2}) become $\left(  2w\partial_{w}-\frac{1}{2}\partial_{u}+\frac
{d-2}{2}\right)  \Omega\left(  w,u,x^{\mu}\right)  =0,$ and similarly for
$S_{i}.$ Their general solution for any $w$ is%
\begin{equation}
\Omega\left(  X\right)  =e^{\left(  d-2\right)  u}\phi\left(  x,we^{4u}%
\right)  ,\;S_{i}\left(  X\right)  =e^{\left(  d-2\right)  u}s_{i}\left(
x,we^{4u}\right)  , \label{omegasi}%
\end{equation}
where, except for the overall factors of $e^{\left(  d-2\right)  u},$ the
fields $\phi\left(  x,we^{4u}\right)  $, $s_{i}\left(  x,we^{4u}\right)  $ are
general functions of the variables $x^{\mu}$ and the combination $we^{4u}.$

Now we impose the kinematic equation $G_{MN}=\nabla_{M}V_{N}$ in the form of
the homothety condition $\pounds _{V}G^{MN}=-2G^{MN}$ as explained in
(\ref{KinG})%
\begin{equation}
V^{K}\partial_{K}G^{MN}-\partial_{K}V^{M}G^{KN}-\partial_{K}V^{N}%
G^{MK}=-2G^{MN}. \label{kinematic condition}%
\end{equation}
This is already satisfied for the fixed metric components $G^{ww}=4w,$
$G^{wu}=-1,$ $G^{uu}=G^{w\mu}=0,$ while it gives the following conditions on
the remaining metric components%
\begin{equation}
\left(  2w\partial_{w}-\frac{1}{2}\partial_{u}\right)  G^{\mu u}=-2G^{\mu
u},\;\;\left(  2w\partial_{w}-\frac{1}{2}\partial_{u}\right)  G^{\mu\nu
}=-2G^{\mu\nu}.
\end{equation}
Their general solutions are
\begin{equation}
G^{\mu u}\left(  X\right)  =e^{4u}\gamma^{\mu}\left(  x,we^{4u}\right)
,\;\;G^{\mu\nu}\left(  X\right)  =e^{4u}\tilde{g}^{\mu\nu}\left(
x,we^{4u}\right)  ,
\end{equation}
where, $\gamma^{\mu}$ and $\tilde{g}^{\mu\nu}$ are general functions of
$x^{\mu}$ and $we^{4u}.$

We see that the solution to the kinematic conditions are given in terms of
functions of fewer than $d+2$ variables. We find that there are remaining
coordinate transformation symmetries in $d+1$ variables that can remove the
$\gamma^{\mu}\left(  x,we^{4u}\right)  ,$ thus reducing further the degrees of
freedom. To explain this we first examine the coordinate transformations that
maintain the restricted form of $G^{MN}$ that emerged above. The infinitesimal
general coordinate transformation of the scalars $W,\Omega,S_{i}$ and the
metric $G^{MN}$ are%
\begin{gather}
\delta_{\varepsilon}X^{M}=\varepsilon^{M}\left(  X\right)  ,\;\;\delta
_{\varepsilon}W=\varepsilon^{K}\partial_{K}W,\;\;\delta_{\varepsilon}%
\Omega=\varepsilon^{K}\partial_{K}\Omega,\;\;\delta_{\varepsilon}%
S_{i}=\varepsilon^{K}\partial_{K}S_{i},\;\;\label{gen1}\\
\delta_{\varepsilon}G^{MN}=\varepsilon^{K}\partial_{K}G^{MN}-\left(
\partial_{K}\varepsilon^{M}\right)  G^{KN}-\left(  \partial_{K}\varepsilon
^{N}\right)  G^{KM}. \label{gen2}%
\end{gather}
The remaining symmetry should not change the form of $W=w$ and the fixed
metric components $G^{ww},G^{wu},G^{w\mu},G^{uu}$ given above$.$ This
requirement is satisfied by the following form of infinitesimal coordinate
transformations $\varepsilon^{M}\left(  X\right)  $
\begin{equation}
\varepsilon^{w}\left(  X\right)  =0,\;\varepsilon^{u}\left(  X\right)
=\Lambda\left(  x,we^{4u}\right)  ,\;\;\varepsilon^{\mu}\left(  X\right)
=\varepsilon^{\mu}\left(  x,we^{4u}\right)  .
\end{equation}
which give $\delta_{\varepsilon}W=\delta_{\varepsilon}G^{ww}=\delta
_{\varepsilon}G^{wu}=\delta_{\varepsilon}G^{uu}=\delta_{\varepsilon}G^{w\mu
}=0.$ In what follows we will show that $\varepsilon^{\mu}\left(
x,we^{4u}\right)  $ at $w=0$ will be related to coordinate transformations in
the $d$ dimensional shadow, while $\Lambda\left(  x,we^{4u}\right)  $ at
$w=0,$ which comes from coordinate transformations of $u,$ will be related to
local scale transformations in the $d$ dimensional shadow.

The coordinate transformations of $\left(  u,x^{\mu}\right)  $ with parameters
$\Lambda\left(  x,we^{4u}\right)  ,\varepsilon^{\mu}\left(  x,we^{4u}\right)
$ give non-zero $\delta_{\varepsilon}G^{\mu u},\delta_{\varepsilon}G^{\mu\nu
},\delta_{\varepsilon}\Omega,\delta_{\varepsilon}S_{i}$. We focus on
$\delta_{\varepsilon}\Omega$ and $\delta_{\varepsilon}G^{\mu u}$ which follow
from (\ref{gen1},\ref{gen2})%
\begin{align}
\delta_{\varepsilon}\Omega &  =\Lambda\partial_{u}\Omega+\varepsilon^{\lambda
}\partial_{\lambda}\Omega\label{dOm}\\
\delta_{\varepsilon}G^{\mu u}  &  =\left\{
\begin{array}
[c]{l}%
\Lambda\partial_{u}G^{\mu u}+\varepsilon^{\lambda}\partial_{\lambda}G^{\mu
u}+\left(  \partial_{w}\varepsilon^{\mu}\right) \\
-\left(  \partial_{\lambda}\varepsilon^{\mu}\right)  G^{\lambda u}-\left(
\partial_{\lambda}\Lambda\right)  G^{\mu\lambda}-\left(  \partial_{u}%
\Lambda\right)  G^{\mu u}%
\end{array}
\right\}  \label{dGum}%
\end{align}
Evidently there is enough gauge freedom in $\Lambda\left(  x,we^{4u}\right)  $
to gauge fix $\Omega=e^{\left(  d-2\right)  u}\phi\left(  x,we^{4u}\right)  $
completely to any desired form as a function of $\left(  x,we^{4u}\right)  $.
We will take advantage of this freedom later\footnote{Convenient gauges will
be mentioned in discussing Eqs.(\ref{trg1}-\ref{g1mn}). We mention here other
possibilities that may serve different purposes. One possible partial gauge
choice is to make $\Omega$ independent of $w,$ as $\Omega=e^{\left(
d-2\right)  u}\phi\left(  x\right)  ,$ while $S_{i}$ remains as given in
(\ref{omegasi}). With this there still remains the gauge freedom of making
$\phi\left(  x\right)  $ a constant. Another gauge of interest is to fix
$\Omega$ such that $\Omega^{2}-aS_{i}^{2}=e^{2\left(  d-2\right)  u}\left[
\phi^{2}\left(  x\right)  -s_{i}^{2}\left(  x\right)  \right]  ,$ is
independent of $w,$ where $\phi\left(  x\right)  ,s_{i}\left(  x\right)  $ are
the shadows defined by the expansions in Eqs.(\ref{ome},\ref{sii}), and again
the $x$ dependence can be further gauge fixed to a constant. This is similar
to Eq.(\ref{nonlinear}), but includes the $u,w$ dependence, thus providing the
prolongation of the shadow for Eq.(\ref{nonlinear})$.$ The expansion in powers
of $w$ in the second gauge gives the details of how the prolongations are
gauge fixed, namely $\phi\phi_{1}-as_{i}s_{1i}=0$ and $\phi\phi_{2}%
-as_{i}s_{2i}+\phi_{1}^{2}-as_{1i}^{2}=0,$ etc., (rather than $\phi_{1}%
=\phi_{2}=0,$ etc., in the first gauge) where $\phi_{1},\phi_{2},s_{1i}%
,s_{2i}$ are defined in Eqs.(\ref{ome},\ref{sii}). \label{fullWeyl}}.

Similarly, there is enough gauge freedom in $\varepsilon^{\mu}\left(
x,we^{4u}\right)  $ to gauge fix $G^{\mu u}=e^{4u}\gamma^{\mu}\left(
x,we^{4u}\right)  =0.$ Then the gauge fixed form of $G^{MN}$ for any $w$
becomes%
\begin{equation}
G^{MN}=%
\begin{array}
[c]{cc}%
M\backslash N &
\begin{array}
[c]{ccc}%
w & \text{~}u & \text{ \ }\nu\;\;\;\;\;\ \;\;\;\;\;\;\;
\end{array}
\\%
\begin{array}
[c]{c}%
w\\
u\\
\mu
\end{array}
& \left(
\begin{array}
[c]{ccc}%
4w & -1 & 0\\
-1 & 0 & 0\\
0 & 0 & e^{4u}\tilde{g}^{\mu\nu}\left(  x,we^{4u}\right)
\end{array}
\right)  .
\end{array}
\label{g-1}%
\end{equation}
The metric with lower indices is then%
\begin{equation}
G_{MN}=%
\begin{array}
[c]{cc}%
M\backslash N &
\begin{array}
[c]{ccc}%
w & ~~\text{~}u~ & \text{ \ }\nu\;\;\;\;\;\ \;\;\;\;\;\;\;
\end{array}
\\%
\begin{array}
[c]{c}%
w\\
u\\
\mu
\end{array}
& \left(
\begin{array}
[c]{ccc}%
0 & -1 & 0\\
-1 & -4w & 0\\
0 & 0 & e^{-4u}\tilde{g}_{\mu\nu}\left(  x,we^{4u}\right)
\end{array}
\right)  .
\end{array}
\label{g}%
\end{equation}
We may now ask if there is any more remaining symmetry that does not change
the gauge fixed forms of $G_{MN}$? For keeping the form of $G_{MN}$ we need
$\delta_{\varepsilon}G^{\mu u}=0$ for the expressions in Eq.(\ref{dGum}) after
setting $G^{\mu u}=0.$ This is satisfied by parameters that obey the condition
$\partial_{w}\varepsilon^{\mu}=G^{\mu\nu}\partial_{\nu}\Lambda,$ with an
arbitrary $\Lambda\left(  x,we^{4u}\right)  .$ To analyze further the meaning
of the remaining symmetry we expand in powers of $w$
\begin{align}
\delta x^{\mu}  &  =\varepsilon^{\mu}\left(  x,we^{4u}\right)  =\varepsilon
_{0}^{\mu}\left(  x\right)  +we^{4u}\varepsilon_{1}^{\mu}\left(  x\right)
+\cdots\label{par1}\\
\delta u  &  =\Lambda\left(  x,we^{4u}\right)  =\Lambda_{0}\left(  x\right)
+we^{4u}\Lambda_{1}\left(  x\right)  +\cdots\label{par2}%
\end{align}
The remaining symmetry has as independent parameters only the lowest component
$\varepsilon_{0}^{\mu}\left(  x\right)  ,$and all $\Lambda\left(
x,we^{4u}\right)  $%
\begin{equation}%
\begin{array}
[c]{l}%
\text{independent:\ }\varepsilon_{0}^{\mu}\left(  x\right)  ,\text{ and
}\Lambda_{0}\left(  x\right)  ,\Lambda_{1}\left(  x\right)  ,\Lambda
_{2}\left(  x\right)  ,\cdots\\
\text{dependent:\ }\varepsilon_{1}^{\mu}\left(  x\right)  =g^{\mu\nu}%
\partial_{\nu}\Lambda_{0},\;\varepsilon_{2}^{\mu}\left(  x\right)  =g^{\mu\nu
}\partial_{\nu}\Lambda_{1}-g_{1}^{\mu\nu}\partial_{\nu}\Lambda_{0}%
,~\text{etc.,}%
\end{array}
\label{e1L1}%
\end{equation}
where $g^{\mu\nu},g_{1}^{\mu\nu}$ are defined by the expansion of the metric
in powers of $we^{4u}$ given below in Eq.(\ref{g-1123}). Among these,
$\varepsilon_{0}^{\mu}\left(  x\right)  $ corresponds to general coordinate
transformations of $x^{\mu}$ while $\Lambda_{0}\left(  x\right)  $ is the
gauge parameter of local scale transformations on the remaining local fields,
known as the Weyl transformations in 1T field theory, as explained below.

The remaining gauge parameters $\Lambda_{n\geq1}\left(  x\right)  $ are
generalizations of the Weyl symmetry $\Lambda_{0}\left(  x\right)  .$ They can
be used to make convenient gauge choices$^{\text{\ref{fullWeyl}}}$.

The transformation of the scalars in (\ref{omegasi}) and metric components in
(\ref{g}) under the remaining symmetry (\ref{par1},\ref{par2}) can be
extracted from the general coordinate transformation rules (\ref{gen1}%
,\ref{gen2}) in the form
\begin{align}
\delta\phi\left(  x,we^{4u}\right)   &  =\left[  \Lambda\left(  x,we^{4u}%
\right)  \left(  4w\partial_{w}+d-2\right)  +\varepsilon^{\mu}\left(
x,we^{4u}\right)  \partial_{\mu}\right]  \phi\left(  x,we^{4u}\right)
,\label{df}\\
\delta s_{i}\left(  x,we^{4u}\right)   &  =\left[  \Lambda\left(
x,we^{4u}\right)  \left(  4w\partial_{w}+d-2\right)  +\varepsilon^{\mu}\left(
x,we^{4u}\right)  \partial_{\mu}\right]  s_{i}\left(  x,we^{4u}\right)
,\label{dsi}\\
\delta\tilde{g}_{\mu\nu}\left(  x,we^{4u}\right)   &  =\Lambda\left(
x,we^{4u}\right)  \left(  4w\partial_{w}-4\right)  \tilde{g}_{\mu\nu}\left(
x,we^{4u}\right)  +\pounds _{\varepsilon}\tilde{g}_{\mu\nu}\left(
x,we^{4u}\right)  . \label{dgmn}%
\end{align}
where $\pounds _{\varepsilon}\tilde{g}^{\mu\nu}\left(  x,we^{4u}\right)  $ is
the Lie derivative using the vector $\varepsilon^{\mu}\left(  x,we^{4u}%
\right)  .$ After inserting in these expressions the field configurations
(\ref{omegasi}-\ref{g}) and the form of the remaining parameters
(\ref{par1},\ref{par2}), the result can be expanded in powers of $w$ to
extract term by term the transformation properties of the shadows in $x^{\mu}$
and their prolongations into the $u$ and $w$ dimensions$.$ To do this we
expand every field in powers of $w$ to define the shadow fields in $d$
dimensions $\phi\left(  x\right)  ,s_{i}\left(  x\right)  ,g_{\mu\nu}\left(
x\right)  $ as the zeroth order terms, while their prolongations $\phi
_{n}\left(  x\right)  ,s_{ni}\left(  x\right)  ,g_{n\mu\nu}\left(  x\right)  $
are defined as the coefficients of the higher powers of $we^{4u}$ as follows
\begin{align}
\phi\left(  x,we^{4u}\right)   &  =\phi\left(  x\right)  +we^{4u}\phi
_{1}\left(  x\right)  +\frac{1}{2}\left(  we^{4u}\right)  ^{2}\phi_{2}\left(
x\right)  +\cdots,\label{ome}\\
s_{i}\left(  x,we^{4u}\right)   &  =s_{i}\left(  x\right)  +we^{4u}%
s_{1i}\left(  x\right)  +\frac{1}{2}\left(  we^{4u}\right)  ^{2}s_{2i}\left(
x\right)  +\cdots, \label{sii}%
\end{align}
Similarly we have for the metric
\begin{equation}
\tilde{g}_{\mu\nu}\left(  x,we^{4u}\right)  =g_{\mu\nu}\left(  x\right)
+we^{4u}g_{1\mu\nu}\left(  x\right)  +\frac{1}{2}\left(  we^{4u}\right)
^{2}g_{2\mu\nu}\left(  x\right)  +\cdots. \label{g123}%
\end{equation}
For the determinant we get%
\begin{equation}
\sqrt{G}=\left\{
\begin{array}
[c]{l}%
=e^{-2du}\sqrt{-\tilde{g}\left(  x,we^{4u}\right)  }\\
=e^{-2du}\sqrt{-g}\left[  1+\frac{we^{4u}}{2}g_{1\lambda}^{\lambda}%
+\frac{(we^{4u})^{2}}{4}\left(  g_{2\lambda}^{\lambda}+\left(  g_{1\lambda
}^{\lambda}\right)  ^{2}\right)  +\cdots\right]  .
\end{array}
\right\}  \label{det}%
\end{equation}
The inverse metric is also computed in terms of $g_{\mu\nu},g_{1\mu\nu
},g_{2\mu\nu},\cdots$ as
\begin{equation}
\tilde{g}^{\mu\nu}\left(  x,we^{4u}\right)  =g^{\mu\nu}\left(  x\right)
-we^{4u}g_{1}^{\mu\nu}\left(  x\right)  -\frac{1}{2}\left(  we^{4u}\right)
^{2}\left(  g_{2}^{\mu\nu}-2g_{1\sigma}^{\nu}g_{1}^{\sigma\mu}\right)  \left(
x\right)  +\cdots. \label{g-1123}%
\end{equation}
Here the upper indices on $g_{1}^{\mu\nu},g_{2}^{\mu\nu},$ etc. are raised or
lowered by using the lowest component of the metric $g_{\mu\nu};$ so
$g_{1}^{\mu\nu},g_{2}^{\mu\nu}$ do not mean the inverses of $g_{1\mu\nu
},g_{2\mu\nu}.$ Inserting these expressions allows us to extract the following
transformation rules for the shadow fields $\phi\left(  x\right)
,s_{i}\left(  x\right)  ,g_{\mu\nu}\left(  x\right)  $ by setting $w=0$ in
Eqs.(\ref{df}-\ref{dgmn})
\begin{align}
\delta\phi\left(  x\right)   &  =\left(  d-2\right)  \Lambda_{0}\left(
x\right)  \phi\left(  x\right)  +\varepsilon_{0}^{\mu}\left(  x\right)
\partial_{\mu}\phi\left(  x\right)  ,\label{ga0}\\
\delta s_{i}\left(  x\right)   &  =\left(  d-2\right)  \Lambda_{0}\left(
x\right)  s_{i}\left(  x\right)  +\varepsilon_{0}^{\mu}\left(  x\right)
\partial_{\mu}s_{i}\left(  x\right)  ,\\
\delta g_{\mu\nu}\left(  x\right)   &  =-4\Lambda_{0}\left(  x\right)
g_{\mu\nu}\left(  x\right)  +\pounds _{\varepsilon_{0}}g_{\mu\nu}\left(
x\right)  . \label{ga03}%
\end{align}
In these expressions it is clear that $\Lambda_{0}\left(  x\right)  $ is the
infinitesimal parameter of the Weyl transformations, which is seen by
comparing to Eq.(\ref{rescale}) and setting $\Lambda_{0}\left(  x\right)
=-\lambda\left(  x\right)  /2$. This shows that the local scale symmetry in 1T
field theory comes from the coordinate reparametrization symmetry $\delta
u=\left[  \Lambda\left(  x,we^{4u}\right)  \right]  _{w=0}$ of the 2T field
theory. This was one of the points we wanted to prove in this section.

The higher powers in $w$ of Eqs.(\ref{df}-\ref{dgmn}) give the nontrivial
transformation rules for the prolongations under coordinate, Weyl and
generalized Weyl transformations $\varepsilon_{0}^{\mu}\left(  x\right)
,\Lambda_{0}\left(  x\right)  ,\Lambda_{n\geq1}\left(  x\right)  $, as follows%
\begin{align}
\delta\phi_{1}\left(  x\right)   &  =\left\{
\begin{array}
[c]{l}%
\left(  d+2\right)  \Lambda_{0}\phi_{1}\left(  x\right)  +\left(  d-2\right)
\Lambda_{1}\phi\\
+\varepsilon_{0}^{\mu}\partial_{\mu}\phi_{1}+\varepsilon_{1}^{\mu}\left(
x\right)  \partial_{\mu}\phi
\end{array}
\right\}  ,\label{gaf1}\\
\delta\phi_{2}\left(  x\right)   &  =\left\{
\begin{array}
[c]{l}%
\left(  d+6\right)  \Lambda_{0}\phi_{2}+2\left(  d+2\right)  \Lambda_{1}%
\phi_{1}+\left(  d-2\right)  \Lambda_{2}\phi\\
+\varepsilon_{0}^{\mu}\partial_{\mu}\phi_{2}+2\varepsilon_{1}^{\mu}%
\partial_{\mu}\phi_{1}+\varepsilon_{2}^{\mu}\partial_{\mu}\phi
\end{array}
\right\}  . \label{gaf2}%
\end{align}
and similarly for $\delta s_{ni}.$ Evidently the terms containing $\Lambda
_{0}\left(  x\right)  $ and $\varepsilon_{0}^{\mu}$ are the local scale
transformations and local coordinate transformations on these fields. Recall
that $\varepsilon_{n\geq1}^{\mu}$ are functions of the $\Lambda_{n}$ as given
in (\ref{e1L1}). Similarly, for the metric prolongations we get the following
transformation laws under the coordinate, Weyl and generalized Weyl
transformations
\begin{align}
\delta g_{1\mu\nu}\left(  x\right)   &  =\left\{
\begin{array}
[c]{l}%
0\times\Lambda_{0}\left(  x\right)  g_{1\mu\nu}\left(  x\right)  +4\Lambda
_{1}\left(  x\right)  g_{\mu\nu}\\
+\pounds _{\varepsilon_{0}}g_{1\mu\nu}\left(  x\right)  +\pounds _{\varepsilon
_{1}}g_{\mu\nu}\left(  x\right)
\end{array}
\right\}  ,\label{gag1}\\
\delta g_{2\mu\nu}\left(  x\right)   &  =\left\{
\begin{array}
[c]{c}%
4\Lambda_{0}\left(  x\right)  g_{2\mu\nu}\left(  x\right)  +8\Lambda
_{1}\left(  x\right)  g_{1\mu\nu}+4\Lambda_{2}g_{\mu\nu}\\
+\pounds _{\varepsilon_{0}}g_{2\mu\nu}\left(  x\right)
+2\pounds _{\varepsilon_{1}}g_{1\mu\nu}\left(  x\right)
+\pounds _{\varepsilon_{2}}g_{\mu\nu}\left(  x\right)
\end{array}
\right\}  . \label{gag2}%
\end{align}

\section{Riemann and Lorentz curvatures}

\label{curve}

We are now ready to use the gauge fixed metric in Eqs.(\ref{g-1}%
,\ref{g},\ref{g123},\ref{g-1123}) to compute the curvatures at any $w$. For
the Christoffel connection $\Gamma_{MN}^{P}\equiv\frac{1}{2}G^{PQ}\left(
\partial_{M}G_{NQ}+\partial_{N}G_{MQ}-\partial_{Q}G_{MN}\right)  $ we obtain%
\begin{align}
\Gamma_{MN}^{w}  &  =%
\begin{array}
[c]{cc}%
M\backslash N &
\begin{array}
[c]{ccc}%
w & \;~u\; & \;\;\;\;\nu\;\;\;\;~~
\end{array}
\\%
\begin{array}
[c]{c}%
w\\
u\\
\mu
\end{array}
& \left(
\begin{array}
[c]{ccc}%
0 & \;2 & 0\\
2 & \;\;8w & 0\\
0 & \;0 & -2e^{-4u}\tilde{g}_{\mu\nu}%
\end{array}
\right)
\end{array}
,\;\Gamma_{MN}^{u}=%
\begin{array}
[c]{cc}%
M\backslash N &
\begin{array}
[c]{ccc}%
\;w\; & ~\;u\;\; & \;\;\;\nu\;\;\;\;\;\;~
\end{array}
\\%
\begin{array}
[c]{c}%
w\\
u\\
\mu
\end{array}
& \left(
\begin{array}
[c]{ccc}%
0 & \;\;0 & 0\\
0 & \;-2\; & 0\\
0 & \;\;\;0 & \frac{e^{-4u}}{2}\partial_{w}\tilde{g}_{\mu\nu}%
\end{array}
\right)
\end{array}
\label{gammawu}\\
\Gamma_{MN}^{\lambda}  &  =%
\begin{array}
[c]{cc}%
M\backslash N &
\begin{array}
[c]{ccc}%
\;\;\;w\;\;\;\;\; & ~\;\;\;\;\;\;\;\;\;\;u\;\;\;\;\;\;\;\;\;\;\; &
\;\;\text{\ \ }\;\;\;\nu\;\;\;\;~~~\;\;\;\;\;\;\;\;\;
\end{array}
\\%
\begin{array}
[c]{c}%
w\\
u\\
\mu
\end{array}
& \left(
\begin{array}
[c]{ccc}%
0 & 0 & \frac{1}{2}\tilde{g}^{\lambda\sigma}\partial_{w}\tilde{g}_{\sigma\nu
}\\
0 & 0 & -2\delta_{\nu}^{\lambda}+2w\tilde{g}^{\lambda\sigma}\partial_{w}%
\tilde{g}_{\sigma\nu}\\
\frac{1}{2}\tilde{g}^{\lambda\sigma}\partial_{w}\tilde{g}_{\sigma\mu} &
-2\delta_{\mu}^{\lambda}+2w\tilde{g}^{\lambda\sigma}\partial_{w}\tilde
{g}_{\sigma\mu} & \Gamma_{\rho\sigma}^{\mu}\left(  \tilde{g}\right)
\end{array}
\right)
\end{array}
\label{gammamu}%
\end{align}
Expanding $\Gamma_{\rho\sigma}^{\mu}\left(  \tilde{g}\right)  $ in the last
line in powers of $w$ gives $\Gamma_{\rho\sigma}^{\mu}\left(  \tilde
{g}\right)  =\Gamma_{\rho\sigma}^{\mu}\left(  g\right)  +we^{4u}\Gamma
_{1\rho\sigma}^{\mu}+\cdots,$ where the zeroth order term is the usual
$\Gamma_{\rho\sigma}^{\mu}\left(  g\right)  $ in $d$ dimensions and the first
order term is%

\begin{equation}
\Gamma_{1\rho\sigma}^{\mu}=\left\{
\begin{array}
[c]{c}%
-\frac{1}{2}g_{1}^{\mu\nu}\left(  \partial_{\rho}g_{\sigma\nu}+\partial
_{\sigma}g_{\rho\nu}-\partial_{\nu}g_{\rho\sigma}\right) \\
+\frac{1}{2}g^{\mu\nu}\left(  \partial_{\rho}g_{1\sigma\nu}+\partial_{\sigma
}g_{1\rho\nu}-\partial_{\nu}g_{1\rho\sigma}\right)
\end{array}
\right\}  . \label{Gam1}%
\end{equation}
Even though $w$ is set to zero eventually, one must first take derivatives of
$\Gamma_{MN}^{P}$ with respect to $w$ in computing various components of the
curvature $R_{MNPQ}\left(  G\right)  .$ Therefore $w$ dependent terms in
$\Gamma_{MN}^{P}$ (i.e. prolongations of its shadow) will contribute to the
curvature in zeroth order in powers of $w$ because of derivatives with respect
to $w.$

\ \ \ To calculate the Riemann tensor $R_{PMN}^{Q}\equiv\partial_{M}%
\Gamma_{NP}^{Q}-\partial_{N}\Gamma_{MP}^{Q}+\Gamma_{MS}^{Q}\Gamma_{NP}%
^{S}-\Gamma_{NS}^{Q}\Gamma_{MP}^{S}$ , we recall that the zero torsion
condition imposes the following kinematical constraint (\ref{RmnpqSymm}) on
the curvature%
\begin{equation}
V_{Q}R_{PMN}^{Q}=V^{Q}R_{QPMN}=V^{Q}R_{MNPQ}=V^{Q}R_{NPQ}^{M}=0.
\label{torsionless}%
\end{equation}
With the gauge choice of Eq.(\ref{gaugeW}) these conditions become%
\begin{equation}
R_{PMN}^{w}=0,\;R_{uPMN}=4wR_{wPMN},\;\;R_{MNu}^{P}=4wR_{MNw}^{P}.
\end{equation}
From the form of the gauge fixed metric in Eq.(\ref{g-1}) we also obtain
\begin{equation}
R_{PMN}^{u}=-R_{wPMN}.
\end{equation}
From these it is easy to see consequences such as
\begin{align}
R_{wMN}^{u}  &  =R_{uMN}^{u}=R_{uwMN}=R_{MNuw}=0,\;\;\\
R_{MNw}^{u}  &  =R_{NMw}^{u},\;\;R_{MNu}^{\lambda}=4wR_{MNw}^{\lambda},\;etc.
\end{align}
Using the antisymmetry and cyclic properties in (\ref{RmnpqSymm}), these
kinematic relations explain many of the results in the following lists for the
Riemann tensor computed by using the Christoffel connection in (\ref{gammawu}%
,\ref{gammamu}) at any $w$
\begin{align}
R_{MPN}^{w}  &  =0,\\
R_{PMN}^{u}  &  =\left\{
\begin{tabular}
[c]{ll}%
$R_{wMN}^{u}=R_{uMN}^{u}=0,\;\;\;\;$ & $R_{\rho\mu\nu}^{u}=\frac{1}{2}%
\nabla_{\lbrack\mu}g_{1\nu]\rho}+\cdots,$\\
$R_{\rho\mu u}^{u}=R_{\mu\rho u}^{u}=4wR_{\rho\mu w}^{u},\;\;$ & $R_{\rho\mu
w}^{u}=\frac{e^{4u}}{4}\left(  g_{1\mu}^{\sigma}g_{1\sigma\rho}-2g_{2\rho\mu
}\right)  +\cdots,$%
\end{tabular}
\right.
\end{align}
where the covariant derivative $\nabla_{\mu}$ is with respect to the metric
$g_{\mu\nu}\left(  x\right)  .$ The curvatures on the first column are either
identically zero or vanish when $w=0,$ while those in the second column
$R_{\rho\mu w}^{u},R_{\rho\mu\nu}^{u}$ do not vanish even at $w=0$. The
$``+\cdots"$ means there are terms proportional to higher powers of $w$ but
are of no interest in our analysis. Similarly we obtain $R_{PMN}^{\lambda}$
with analogous properties for the first and second columns
\begin{equation}
R_{PMN}^{\lambda}:\left\{
\begin{tabular}
[c]{ll}%
$R_{P\mu u}^{\lambda}=4wR_{P\mu w}^{\lambda},$ & $R_{Pwu}^{\lambda}=0,$\\
$R_{w\mu u}^{\lambda}=R_{u\mu w}^{\lambda}=4wR_{w\mu w}^{\lambda},$ & $R_{w\mu
w}^{\lambda}=\frac{e^{8u}}{4}\left(  g_{1}^{\lambda\nu}g_{1\mu\nu}-2g_{2\mu
}^{\lambda}\right)  +\cdots,$\\
$R_{u\mu u}^{\lambda}=16w^{2}R_{w\mu w}^{\lambda},\;\;$ & $R_{w\mu\nu
}^{\lambda}=\frac{e^{4u}}{2}\nabla_{\lbrack\mu}g_{1\nu]}^{\lambda}+\cdots,$\\
$R_{\rho\mu u}^{\lambda}=-R_{\rho u\mu}^{\lambda}=4wR_{\rho\mu w}^{\lambda
},\;\;\;$ & $R_{\rho\mu w}^{\lambda}=-R_{\rho w\mu}^{\lambda}=\frac{e^{4u}}%
{2}g^{\lambda\sigma}\nabla_{\lbrack\sigma}g_{1\rho]\mu}+\cdots$\\
$R_{u\mu\nu}^{\lambda}=4wR_{w\mu\nu}^{\lambda},$ & $R_{\rho\mu\nu}^{\lambda
}=R_{\rho\mu\nu}^{\lambda}\left(  g\right)  -g_{1[\mu}^{\lambda}g_{\nu]\rho
}-\delta_{\lbrack\mu}^{\lambda}g_{1\nu]\rho}+\cdots.$%
\end{tabular}
\right.
\end{equation}
At $w=0$ the nonvanishing components of $R_{QPMN}$ with all lower indices are
$R_{w\mu w\nu}$, $R_{\mu\nu\lambda w}$, $R_{\mu\nu\lambda\sigma}$
\begin{equation}
w\rightarrow0:\left\{
\begin{array}
[c]{l}%
R_{w\mu w\nu}\left(  G\right)  =\frac{e^{4u}}{4}\left(  g_{1\mu\sigma}g_{1\nu
}^{\sigma}-2g_{2\mu\nu}\right)  +\cdots\\
R_{\mu\nu\lambda w}\left(  G\right)  =\frac{1}{2}\left(  \nabla_{\mu}%
g_{1\nu\lambda}-\nabla_{\nu}g_{1\mu\lambda}\right)  +\cdots\\
R_{\mu\nu\lambda\sigma}\left(  G\right)  =e^{-4u}\left[  R_{\mu\nu
\lambda\sigma}\left(  g\right)  +g_{1\sigma\lbrack\mu}g_{\nu]\lambda
}-g_{1\lambda\lbrack\mu}g_{\nu]\sigma}\right]  +\cdots
\end{array}
\right.  \; \label{curves}%
\end{equation}
In the last expression it should be noted that $R_{\rho\mu\nu}^{\lambda
}\left(  G\right)  $ differs from $R_{\rho\mu\nu}^{\lambda}\left(  g\right)
$, the latter being the standard Riemann tensor constructed from the metric
$g_{\mu\nu}.$ The difference is accounted by the contributions of the
prolongations of the metric which contribute to $R_{\rho\mu\nu}^{\lambda
}\left(  G\right)  $ even when $w=0.$

We can now compute the Ricci tensor $R_{MN}\equiv R_{MPN}^{P}=R_{MwN}%
^{w}+R_{MuN}^{u}+R_{M\lambda N}^{\lambda}.$ The kinematic constraints
$V^{M}R_{MN}=0,$ imply%
\begin{equation}
R_{uN}=4wR_{wN}.
\end{equation}
Hence $R_{uw}$, $R_{u\nu}$ are related to $R_{ww}$, $R_{w\nu}$ respectively by
a factor of $4w$ while $R_{uu}=\left(  4w\right)  ^{2}R_{ww},$ hence we have%
\begin{equation}
R_{MN}\left(  G\right)  =%
\begin{array}
[c]{cc}%
M\backslash N &
\begin{array}
[c]{ccc}%
w\;\;\;\; & \text{ \ \ \ }u\text{ \ \ \ \ } & \;\;\text{ \ }\nu\;\;\;
\end{array}
\\%
\begin{array}
[c]{c}%
w\\
u\\
\mu
\end{array}
& \left(
\begin{array}
[c]{ccc}%
R_{ww} & 4wR_{ww} & R_{w\nu}\\
4wR_{ww} & \left(  4w\right)  ^{2}R_{ww} & 4wR_{w\nu}\\
R_{w\mu} & 4wR_{w\mu} & R_{\mu\nu}\left(  G\right)
\end{array}
\right)
\end{array}
\label{RMN}%
\end{equation}
where%
\begin{equation}%
\begin{array}
[c]{l}%
R_{ww}\left(  G\right)  =\frac{e^{8u}}{4}Tr\left(  g_{1}g_{1}-2g_{2}\right)
+\cdots\\
R_{w\mu}\left(  G\right)  =\frac{e^{4u}}{2}\left(  \nabla_{\lambda}g_{1\mu
}^{\lambda}-\nabla_{\mu}Trg_{1}\right)  +\cdots\\
R_{\mu\nu}\left(  G\right)  =R_{\mu\nu}\left(  g\right)  -\left(  d-2\right)
g_{1\mu\nu}-\left(  Trg_{1}\right)  g_{\mu\nu}+\cdots
\end{array}
\label{RMN2}%
\end{equation}
The trace notation $Tr$ means that indices are contracted by using the lowest
mode $g_{\mu\nu}.$ The \textquotedblleft$+\cdots$\textquotedblright\ indicates
that there are additional higher order terms in powers of $w$ that are not of
interest in our analysis. For $w=0$ only $R_{ww},R_{w\mu}$ and $R_{\mu\nu}$
have non-vanishing contributions while the other components of $R_{MN}$
vanish. In the last expression we see that $R_{\mu\nu}\left(  G\right)  $
differs from $R_{\mu\nu}\left(  g\right)  $ which is the standard Ricci tensor
constructed from the metric $g_{\mu\nu}.$

Finally the Ricci scalar, $R\left(  G\right)  =G^{MN}R_{MN}=4wR_{ww}%
-2R_{wu}+e^{4u}\tilde{g}^{\mu\nu}R_{\mu\nu}\left(  G\right)  ,$ is%
\begin{equation}
R\left(  G\right)  =e^{4u}\left[  R\left(  g\right)  -2\left(  d-1\right)
e^{4u}Trg_{1}\right]  +\cdots\label{RG}%
\end{equation}
Again in the last expression $R\left(  G\right)  $ differs from $R\left(
g\right)  $ which is the standard curvature scalar.

As seen explicitly in all the expressions above, the prolongations of the
shadow of the metric, namely $g_{1\mu\nu},g_{2\mu\nu}$ contribute
non-trivially to the prolongations of the curvatures. Even when $w=0,$ there
are non-vanishing curvature components, such as $R_{w\mu w}^{\lambda}%
,R_{w\mu\nu}^{\lambda},R_{\rho\mu w}^{\lambda},R_{\rho\mu\nu}^{\lambda}$ that
point not only in the $x^{\mu}$ directions but also in the $w,u$ directions.
The notation, $R_{\rho\mu\nu}^{\lambda}\left(  G\right)  ,R_{\mu\nu}\left(
G\right)  ,R\left(  G\right)  $ is used to distinguish them from $R_{\rho
\mu\nu}^{\lambda}\left(  g\right)  ,R_{\mu\nu}\left(  g\right)  ,R\left(
g\right)  $ where the latter depend only the lowest mode $g_{\mu\nu}\left(
x\right)  $ while the former depend on $G_{\mu\nu}$ including the higher modes
$g_{1\mu\nu},g_{2\mu\nu}.$ We will see however, that after taking into account
the dynamical equations of motion, all extra curvature pieces get determined
only in terms of the shadow fields $g_{\mu\nu}\left(  x\right)  ,\phi\left(
x\right)  ,s_{i}\left(  x\right)  ,$ while the dynamics of these lowest modes
interacting with $R_{\rho\mu\nu}^{\lambda}\left(  g\right)  ,R_{\mu\nu}\left(
g\right)  ,R\left(  g\right)  $ will be given by standard General Relativity
(with the Weyl symmetry) as determined self consistently only by the shadow
action in Eq.(\ref{S0shad2}).

\subsection{Gauge fixed vielbein, spin connection and SO$\left(  d,2\right)  $
curvature}

\label{vielb}

For completeness we record here also the gauge fixed forms of the vielbein,
spin connection and SO$\left(  d,2\right)  $ curvature $E_{M}^{a}\left(
w,u,x\right)  ,$ $\omega_{M}^{ab}\left(  w,u,x\right)  $, $R_{MN}^{ab}\left(
w,u,x\right)  $ that are compatible with the gauge fixed metric and its
curvatures above.

We take the following form of the gauge fixed vielbein that satisfies
$G_{MN}=E_{M}^{a}E_{N}^{b}\eta_{ab}$ up to a local SO$\left(  d,2\right)  $
transformation in tangent space%

\begin{equation}
E_{M}^{~~a}=%
\begin{array}
[c]{cc}%
M\backslash a &
\begin{array}
[c]{ccc}%
~-^{\prime} & \;+^{\prime} & ~~~\ ~i~~\ \ \ \ \ \ \ \ \ \ \
\end{array}
\;\;\;\\%
\begin{array}
[c]{c}%
w\\
u\\
\mu
\end{array}
& \left(
\begin{array}
[c]{ccc}%
~1~\;\; & \;\;0~\; & 0\\
2w\;\; & 1 & 0\\
0\;\; & 0 & e^{-2u}\tilde{e}_{\mu}^{~i}\left(  x,we^{4u}\right)
\end{array}
\right)
\end{array}
\end{equation}
Its inverse that satisfies $E_{a}^{M}E_{M}^{b}=\delta_{a}^{b}$ or $E_{a}%
^{M}E_{N}^{a}=\delta_{N}^{M}$ is%

\begin{equation}
E_{a}^{~M}=%
\begin{array}
[c]{cc}%
a\backslash M &
\begin{array}
[c]{ccc}%
\;\;w & \;\;~u~ & ~\ \ \ \ ~\mu~~~\ \ \;\;~
\end{array}
\;\;\;\\%
\begin{array}
[c]{c}%
-^{\prime}\\
+^{\prime}\\
i
\end{array}
& \left(
\begin{array}
[c]{ccc}%
~1~\;\; & \;0~\; & 0\\
-2w\;\; & 1 & 0\\
0\; & 0 & e^{2u}\tilde{e}_{i}^{~\mu}\left(  x,we^{4u}\right)
\end{array}
\right)
\end{array}
\;
\end{equation}
where $\tilde{e}_{\mu}^{~i}$ and $\tilde{e}_{i}^{~\mu}$ are inverses of each
other. These may be expanded in powers of $we^{4u}$
\begin{align}
\tilde{e}_{\mu}^{~i}\left(  x,we^{4u}\right)   &  =e_{\mu}^{~i}+we^{4u}%
e_{1\mu}^{i}+\frac{1}{2}\left(  we^{4u}\right)  ^{2}e_{2\mu}^{i}+\cdots\\
\tilde{e}_{i}^{~\mu}\left(  x,we^{4u}\right)   &  =e_{i}^{~\nu}-we^{4u}%
e_{1i}^{\nu}+\frac{\left(  we^{4u}\right)  ^{2}}{2}\left(  2e_{1\rho}^{\nu
}e_{1i}^{\rho}-e_{2i}^{\nu}\right)  +\cdots
\end{align}
Here $e_{i}^{~\nu}$ is the inverse of $e_{\mu}^{~i}$ as usual, but
$e_{1i}^{\mu}$ is not the inverse of $e_{1\mu}^{i}$, rather it is $e_{1\mu
}^{i}$ with indices raised or lowered by using the appropriate tangent space
or base space metrics, $e_{1i}^{\mu}=\eta_{ij}e_{1\nu}^{j}g^{\nu\mu}$, and
similarly for $e_{2i}^{\nu}$. From $\tilde{g}_{\mu\nu}=\tilde{e}_{\mu}%
^{i}\tilde{e}_{\nu}^{i}\eta_{ij}$ we can obtain relations between the
expansion of the vielbein and the expansion of the metric given in (\ref{g123})%

\begin{equation}%
\begin{array}
[c]{l}%
g_{\mu\nu}=e_{\mu}^{~i}e_{\nu}^{~j}\eta_{ij},\\
g_{1\mu\nu}=\left(  e_{1\mu}^{i}e_{\nu}^{~j}+e_{\mu}^{~i}e_{1\nu}^{j}\right)
\eta_{ij},\\
g_{2\mu\nu}=\left(  e_{2\mu}^{i}e_{\nu}^{~j}+e_{\mu}^{~i}e_{2\nu}^{j}%
+2e_{1\mu}^{i}e_{1\nu}^{j}\right)  \eta_{ij}.
\end{array}
\end{equation}

Recall the gauge fixed versions of the vectors $V_{M}=\frac{1}{2}\partial
_{M}W=\left(  \frac{1}{2},0,0\right)  _{M},$ and $V^{M}=\frac{1}{2}%
\partial_{N}WG^{MN}=\left(  2w,-\frac{1}{2},0\right)  ^{M}$ in
Eq.(\ref{gaugeW}). Their tangent space counterparts become $V_{a}=V_{M}%
E_{a}^{M}=\frac{1}{2}E_{a}^{w}$ and $V^{a}=V^{M}E_{M}^{a}=2wE_{w}^{a}-\frac
{1}{2}E_{u}^{a}$. Explicitly these are%
\begin{equation}
V_{a}=\left(  \frac{1}{2},-w,0\right)  _{a},\;V^{a}=\left(  w,-\frac{1}%
{2},0\right)  ^{a},\text{ in the basis }a=\left(  -^{\prime},+^{\prime
},i\right)  , \label{Va}%
\end{equation}
and have the dot product $V^{a}V_{a}=w.$

The spin connection is constructed by using the standard relation $\omega
_{M}^{ab}=E^{Na}E^{Pb}\left(  C_{MNP}-C_{NPM}-C_{PMN}\right)  $ given in
Eqs.(\ref{spinConn}), with $C_{PMN}\equiv-\frac{1}{2}E_{Pa}\left(
\partial_{M}E_{N}^{a}-\partial_{N}E_{M}^{a}\right)  .$ With the above gauge
fixed form of $E_{M}^{a}$ we obtain%
\begin{equation}
\omega_{w}^{ab}=%
\begin{array}
[c]{cc}%
a\backslash b &
\begin{array}
[c]{ccc}%
-^{\prime} & +^{\prime} & \;\;j\;\;\;\;\;\;
\end{array}
\\%
\begin{array}
[c]{c}%
-^{\prime}\\
+^{\prime}\\
i
\end{array}
& \left(
\begin{array}
[c]{ccc}%
\;0\; & \;0\; & 0\\
0 & 0 & 0\\
0 & 0 & \frac{1}{2}\tilde{e}^{\sigma\lbrack i}\partial_{w}\tilde{e}_{\sigma
}^{j]}%
\end{array}
\right)
\end{array}
,\;\;\;\omega_{u}^{ab}=%
\begin{array}
[c]{cc}%
a\backslash b &
\begin{array}
[c]{ccc}%
-^{\prime} & +^{\prime} & \;\;\;j\;\;\;\;\;\;\;\;\;
\end{array}
\\%
\begin{array}
[c]{c}%
-^{\prime}\\
+^{\prime}\\
i
\end{array}
& \left(
\begin{array}
[c]{ccc}%
\;0\; & -2\; & 0\\
2 & 0 & 0\\
0 & 0 & 2w\tilde{e}^{\sigma\lbrack i}\partial_{w}\tilde{e}_{\sigma}^{j]}%
\end{array}
\right)
\end{array}
,\;
\end{equation}
and%
\begin{equation}
\omega_{\lambda}^{ab}=%
\begin{array}
[c]{cc}%
a\backslash b &
\begin{array}
[c]{ccc}%
\;\;\;\text{\ \ \ \ }-^{\prime}\;\;\;\; & \;\;\;\;\;\;\;\;\;\;\;\;\text{
\ }+^{\prime}\text{ \ \ \ \ \ \ \ \ \ \ \ \ \ \ } & \text{ \ \ \ \ }%
\;j\;\;\;\;\;\;\;\;\;\;\;\;\;\;\;\;\;\;\;
\end{array}
\\%
\begin{array}
[c]{c}%
-^{\prime}\\
+^{\prime}\\
i
\end{array}
& \left(
\begin{array}
[c]{ccc}%
0 & 0 & e^{-2u}\left(  -2e_{\lambda}^{j}+w\tilde{e}^{~j\sigma}\partial
_{w}\tilde{g}_{\lambda\sigma}\right) \\
0 & 0 & \frac{e^{-2u}}{2}\tilde{e}^{~j\sigma}\partial_{w}\tilde{g}%
_{\lambda\sigma}\\
e^{-2u}\left(  2\tilde{e}_{\lambda}^{~i}-w\tilde{e}^{~i\sigma}\partial
_{w}\tilde{g}_{\lambda\sigma}\right)  & -\frac{e^{-2u}}{2}\tilde{e}^{~i\sigma
}\partial_{w}\tilde{g}_{\lambda\sigma} & \omega_{\lambda}^{ij}\left(
\tilde{e}\right)
\end{array}
\right)
\end{array}
\end{equation}
where $\omega_{\lambda}^{ij}\left(  \tilde{e}\right)  $ is the standard spin
connection in $d$ dimensions as constructed from $\tilde{e}_{\lambda}%
^{i}\left(  x,we^{4u}\right)  $ including the prolongations of the shadow
$\tilde{e}_{\lambda}^{i}\left(  x\right)  $.

With these explicit forms, it can be verified that the spin connection
$\omega_{M}^{ab},$ the vielbein $E_{M}^{a}$ and the vector $V^{a}$ satisfy the
kinematic relation
\begin{equation}
E_{M}^{a}=D_{M}V^{a}=\partial_{M}V^{a}+\omega_{M}^{ab}V_{b},
\end{equation}
that is required by 2T-gravity as expected from Eq.(\ref{Emi}). The kinematic
equations have completely fixed all components of $\omega_{M}^{ab}\left(
X\right)  $ in terms of $e_{\lambda}^{j}\left(  x,we^{4u}\right)  $ and
explicit functions of the extra coordinates $w,u.$ When $w=0$ we recognize
that the vielbein in $d$ dimensions $e_{\lambda}^{j}\left(  x\right)  $ is
basically the shadow component $\omega_{\lambda}^{-^{\prime}i}$ of the spin
connection that remains unrestricted as a function of $x^{\mu}$ as far as the
kinematic equations are concerned.

The SO$\left(  d,2\right)  $ curvature is%
\begin{equation}
R_{MN}^{ab}=\partial_{M}\omega_{N}^{ab}-\partial_{N}\omega_{M}^{ab}+\omega
_{M}^{ak}\omega_{Nk}^{\text{ \ \ }b}-\omega_{N}^{\text{ }ak}\omega
_{Mk}^{\text{ \ \ \ }b}=-R_{QMN}^{P}E_{P}^{a}E^{Qb}%
\end{equation}
With the help of the antisymmetry $R_{MN}^{ab}=-R_{NM}^{ab},$ $R_{MN}%
^{ab}=-R_{MN}^{ba},$ and the kinematic relations in Eq.(\ref{L1}),
$R_{uN}^{ab}=4wR_{wN}^{ab},$ $R_{MN}^{a-^{\prime}}=2wR_{MN}^{a+^{\prime}},$
all the non zero components of the curvature are determined as follows%
\begin{equation}%
\begin{array}
[c]{lll}%
R_{w\mu}^{+^{\prime}i}=\frac{e^{6u}}{2}\left(  \frac{1}{2}g_{1}^{\lambda
\sigma}g_{1\mu\sigma}-g_{2\mu}^{\lambda}\right)  e_{\lambda}^{i}+\cdots, &
R_{w\mu}^{-^{\prime}i}=2wR_{w\mu}^{+^{\prime}i},\;\;\; & R_{w\mu}^{ij}%
=\tilde{e}^{\nu i}\tilde{e}_{\rho}^{j}R_{\nu w\mu}^{\rho}\left(  G\right) \\
R_{u\mu}^{+^{\prime}i}=4wR_{w\mu}^{+^{\prime}a}, & R_{u\mu}^{-^{\prime}%
i}=2wR_{u\mu}^{+^{\prime}i}, & R_{u\mu}^{ij}=4wR_{w\mu}^{ij}\\
R_{\mu\nu}^{+^{\prime}i}=\frac{e^{2u}}{2}\left(  \nabla_{\mu}g_{1\nu\lambda
}-\nabla_{\nu}g_{1\mu\lambda}\right)  e^{i\lambda}+\cdots, & R_{\mu\nu
}^{-^{\prime}i}=2wR_{\mu\nu}^{+^{\prime}i}, & R_{\mu\nu}^{ij}=\tilde
{e}^{\sigma i}\tilde{e}_{\rho}^{i}R_{\sigma\mu\nu}^{\rho}\left(  G\right)
\end{array}
\end{equation}
where $R_{\nu w\mu}^{\rho}\left(  G\right)  $ and $R_{\sigma\mu\nu}^{\rho
}\left(  G\right)  $ are given in Eq.(\ref{curves}). These are the curvatures
at any $w$ which include all the prolongations of the shadow into the higher
dimensions. When $w=0,$ the nonzero terms are just $R_{w\mu}^{+^{\prime}i}$,
$R_{\mu\nu}^{+^{\prime}i}$, $R_{w\mu}^{ij}$, $R_{\mu\nu}^{ij}$ while all
others vanish.

It should be noted that even at $w=0$ there are non-trivial components of
curvature pointing in the $w$ direction in base space and in the $+^{\prime}$
direction in tangent space. This is part of the information about the
prolongation of the shadow. In the next section it will be shown that, after
taking the dynamical equations into account, only the shadow fields $e_{\mu
}^{i}\left(  x\right)  ,$ together with matter fields such as $\phi\left(
x\right)  ,s_{i}\left(  x\right)  ,$ determine all curvature components
including the prolongations, while the shadow fields satisfy among themselves
the familiar General Relativity equations (with a Weyl symmetry) which follows
self consistently from the 1T-physics shadow action in Eq.(\ref{S0shad2}).

\section{Dynamics of shadows \& prolongations}

\label{dyn}

Having chosen gauges and solved the kinematic equations in the previous
sections, we are now ready to discuss the matching of geometry to matter
through the dynamical equations derived in section (\ref{eoms}) from the
2T-gravity action (\ref{action}) and Table-1\footnote{We have neglected gauge
fields and spinor fields to keep our analysis simple. The same general
conclusions about the shadows are obtained if all of the fields described in
Table-1, that would be required for the Standard Model coupled to gravity, are
included in the present analysis.}
\begin{equation}%
\begin{array}
[c]{l}%
\left[  R_{MN}\left(  G\right)  -S_{MN}\left(  \Omega,S_{i}\right)  \right]
_{W=0}=0,\\
\left[  \frac{1}{\sqrt{G}}\partial_{M}\left(  \sqrt{G}G^{MN}\partial_{N}%
\Omega\right)  -2a\Omega R\left(  G\right)  +a\partial_{\Omega}V\left(
\Omega,S_{i}\right)  \right]  _{W=0}=0,\\
\left[  \frac{1}{\sqrt{G}}\partial_{M}\left(  \sqrt{G}G^{MN}\partial_{N}%
S_{i}\right)  -2aS_{i}R\left(  G\right)  -\partial_{S_{i}}V\left(
\Omega,S_{i}\right)  \right]  _{W=0}=0,
\end{array}
\label{original}%
\end{equation}
where $S_{MN}$ was obtained in Eq.(\ref{SMN})%
\begin{equation}
S_{MN}\left(  \Omega,S_{i}\right)  \equiv\frac{1}{\left(  \Omega^{2}%
-aS_{i}^{2}\right)  }\left[  -\frac{1}{2a}\partial_{M}\Omega\partial_{N}%
\Omega+\frac{1}{2}\partial_{M}S_{i}\partial_{N}S_{i}+\nabla_{M}\partial
_{N}\left(  \Omega^{2}-aS_{i}^{2}\right)  \right]  . \label{SMN2}%
\end{equation}
Note that these 2T-gravity equations are imposed only at $w=0,$ unlike the
kinematic equations that were solved at all $w$ (see explanation in section
\ref{eoms}). We want to compare these equations in $\left(  d+2\right)  $
dimensions to the equations of motion of General Relativity in $d$ dimensions
\begin{equation}%
\begin{array}
[c]{l}%
R_{\mu\nu}\left(  g\right)  =\frac{1}{\left(  \phi^{2}-as_{i}^{2}\right)
}\left[
\begin{array}
[c]{c}%
-\frac{1}{2a}\partial_{\mu}\phi\partial_{\nu}\phi+\frac{1}{2}\partial_{\mu
}s_{i}\partial_{\nu}s_{i}+\nabla_{\mu}\partial_{\nu}\left(  \phi^{2}%
-as_{i}^{2}\right) \\
+\frac{g_{\mu\nu}}{d-2}\left(  V\left(  \phi,s_{i}\right)  +\nabla^{2}\left(
\phi^{2}-as_{i}^{2}\right)  \right)
\end{array}
\right] \\
\frac{1}{\sqrt{-g}}\partial_{\mu}\left(  \sqrt{-g}g^{\mu\nu}\partial_{\nu}%
\phi\right)  =2a\phi R\left(  g\right)  -a\partial_{\phi}V\left(  \phi
,s_{i}\right)  ,\\
\frac{1}{\sqrt{-g}}\partial_{\mu}\left(  \sqrt{-g}g^{\mu\nu}\partial_{\nu
}s_{i}\right)  =2as_{i}R\left(  g\right)  +\partial_{s_{i}}V\left(  \phi
,s_{i}\right)  ,
\end{array}
\label{shadoweoms}%
\end{equation}
that follow directly from varying the conformal shadow action (\ref{S0shad2})
and using (\ref{Rmn},\ref{Tmunu}).

In comparing the original and the shadow equations, we note that we lose two
dimensions not only in the spacetime $X^{M}\rightarrow x^{\mu}$ but also in
the components of the metric $G_{MN}\left(  X\right)  \rightarrow g_{\mu\nu
}\left(  x\right)  $, and similarly for curvature, gauge fields, spinors,
etc.. Recall also that $R_{\mu\nu}\left(  G\right)  ,R\left(  G\right)  $ are
different than the $R_{\mu\nu}\left(  g\right)  ,R\left(  g\right)  $ that
appear in (\ref{shadoweoms}), as seen in Eqs.(\ref{curves}-\ref{R}). The
differences depend on the prolongations of the metric and the scalars given in
Eqs.(\ref{ome}-\ref{g-1123}). Moreover, additional components of the tensor
$R_{MN}\left(  G\right)  $ are restricted by the original equations
(\ref{original}). So, going from (\ref{original}) to (\ref{shadoweoms}) is not
a naive dimensional reduction. The questions we need to investigate include
the following.

\begin{itemize}
\item[(i)] We recall that the conformal shadow \textit{action} (\ref{S0shad2})
was derived in \cite{2Tgravity} from the 2T-gravity \textit{action}
(\ref{action}) by inserting directly the solution of the kinematic equations
and the gauge fixing discussed above. Can the shadow \textit{equations}
(\ref{shadoweoms}) be derived from the original \textit{equations} of motion
(\ref{original}) rather than from varying the shadow action? Sometimes these
two procedures do not agree, so it is important to verify that they give the
same result.

\item[(ii)] More importantly, are the prolongations additional Kaluza-Klein
type degrees of freedom? What is the dynamics of the prolongations of the
metric $G_{MN}$, curvature $R_{PQMN}\left(  G\right)  $, and scalars
$\Omega,S_{i}$, that survived the gauge fixing and kinematic constraints of
the previous sections, and do their dynamics restrict the dynamics of the
shadow fields $\left(  \phi,s_{i},g_{\mu\nu}\right)  $ beyond the equations of
motion in (\ref{shadoweoms})? If additional restrictions on $\left(
\phi,s_{i},g_{\mu\nu}\right)  $ arise it would imply that the shadow action
(\ref{S0shad2}) misses information that influences the shadow fields.
\end{itemize}

As explained below, the answers are that there are non-trivial prolongations
of the metric, curvature and the scalars, which are however determined only by
the shadows $\left(  \phi,s_{i},g_{\mu\nu}\right)  $. Meanwhile, the shadows
themselves are determined self consistently precisely as dictated by the
shadow action (\ref{S0shad2}) which yielded the General Relativity equations
(\ref{shadoweoms}).

To investigate these questions we insert the expansions in powers of $w$ for
the fields (\ref{ome}-\ref{g-1123}) and for the curvatures (\ref{RMN2}%
,\ref{R}) into the original equations (\ref{original}). The derivatives
$\partial_{w},\partial_{u}$ in the scalar equations give no new information at
$w=0$ because such terms combine to expressions that are proportional to the
kinematic conditions, which are already satisfied for the scalars. This is a
non-trivial result that is true in curved space only for the special value of
$a=\left(  d-2\right)  /8\left(  d-1\right)  .$ Hence, for the scalar
equations, even though the prolongations $\phi_{1},\phi_{2},s_{1i},s_{2i}%
,$etc. are non-zero, we obtain directly the naive reduction of the $d+2$
dimensional equations to $d$ dimensions, in agreement with the shadow
equations (\ref{shadoweoms}). The prolongations of the scalars $\phi_{1}%
,\phi_{2},s_{1i},s_{2i},$etc. are not fixed by the scalar equations in
(\ref{original}).

Turning to the curvature equation, $R_{MN}=S_{MN}$ at $w=0,$ we begin by
computing $\left[  S_{\mu\nu}\left(  \Omega,S_{i}\right)  \right]  _{w=0}$
from (\ref{SMN2}) as follows%
\begin{equation}
\left[  S_{\mu\nu}\left(  \Omega,S_{i}\right)  \right]  _{w=0}=\frac
{1}{\left(  \phi^{2}-as_{i}^{2}\right)  }\left[
\begin{array}
[c]{c}%
-\frac{1}{2a}\partial_{\mu}\phi\partial_{\nu}\phi+\frac{1}{2}\partial_{\mu
}s_{i}\partial_{\nu}s_{i}+\nabla_{\mu}\partial_{\nu}\left(  \phi^{2}%
-as_{i}^{2}\right) \\
-\left\{  \left(  \Gamma_{\mu\nu}^{w}\partial_{w}+\Gamma_{\mu\nu}^{u}%
\partial_{u}\right)  \left(  \Omega^{2}-aS_{i}^{2}\right)  \right\}  _{w=0}%
\end{array}
\right]
\end{equation}
After inserting the explicit Christoffel symbols $\Gamma_{\mu\nu}^{w}%
,\Gamma_{\mu\nu}^{u}$ in Eqs.(\ref{gammawu},\ref{gammamu},\ref{Gam1}) and
setting $w=0,$ we obtain%
\begin{equation}
\left\{  \left(  \Gamma_{\mu\nu}^{w}\partial_{w}+\Gamma_{\mu\nu}^{u}%
\partial_{u}\right)  \left(  \Omega^{2}-aS_{i}^{2}\right)  \right\}
_{w=0}=-4\left(  \phi\phi_{1}-as_{i}s_{1i}\right)  g_{\mu\nu}+\left(
d-2\right)  \left(  \phi^{2}-as_{i}^{2}\right)  g_{1\mu\nu}.
\end{equation}
Now matching geometry with matter $\left[  R_{\mu\nu}\left(  G\right)
-S_{\mu\nu}\left(  \Omega,S_{i}\right)  \right]  _{w=0}=0,$ where the
curvature
\begin{equation}
R_{\mu\nu}\left(  G\right)  =R_{\mu\nu}\left(  g\right)  -\left(  d-2\right)
g_{1\mu\nu}-\left(  Trg_{1}\right)  g_{\mu\nu}+\cdots
\end{equation}
was given in (\ref{RMN2}), we find%
\begin{equation}
R_{\mu\nu}\left(  g\right)  =S_{\mu\nu}\left(  \phi,s\right)  +\left[
4\frac{\phi\phi_{1}-as_{i}s_{1i}}{\phi^{2}-as_{i}^{2}}+Tr\left(  g_{1}\right)
\right]  g_{\mu\nu}.
\end{equation}
This agrees with the shadow equations (\ref{shadoweoms}) only if the term in
brackets satisfies
\begin{equation}
4\frac{\phi\phi_{1}-as_{i}s_{1i}}{\phi^{2}-as_{i}^{2}}+Tr\left(  g_{1}\right)
=\frac{V\left(  \phi,s_{i}\right)  +\nabla^{2}\left(  \phi^{2}-as_{i}%
^{2}\right)  }{\left(  d-2\right)  \left(  \phi^{2}-as_{i}^{2}\right)  }.
\label{trg1}%
\end{equation}
In fact, this relation is exactly correct and can be derived directly from
Eq.(\ref{d2WV}), which was obtained as a consequence of the original equations
of 2T-gravity (\ref{d1}-\ref{d3}).

We have thus shown that all the shadow equations (\ref{shadoweoms}) derived
directly from the shadow action (\ref{S0shad2}) are in exact agreement with
solving directly the original equations of motion (\ref{original}) in $d+2$
dimensions. This answers the concerns raised above in item (i).

There remains to examine the rest of the original equations of motion
(\ref{original}) $R_{MN}=S_{MN}$ at $w=0,$ to determine whether any additional
constraints emerge on the shadow fields or their prolongations. On the
geometry side we see from (\ref{RMN}) that $\left[  R_{uw}=R_{uu}=R_{u\mu
}\right]  _{w=0}=0,$ and also on the matter side we find $\left[
S_{uw}=S_{uu}=S_{u\mu}\right]  _{w=0}=0$ for the special value of $a=\left(
d-1\right)  /8\left(  d-2\right)  .$ Therefore the corresponding equations are
identically satisfied without any conditions on the shadows or the
prolongations. Proceeding further, from the remaining two cases $\left[
R_{ww}\left(  G\right)  -S_{ww}\left(  \Omega,S_{i}\right)  \right]  _{w=0}=0$
and $\left[  R_{w\mu}\left(  G\right)  -S_{w\mu}\left(  \Omega,S_{i}\right)
\right]  _{w=0}=0,$ we get non-trivial equations that restrict the
prolongations%
\begin{equation}
Tr\left(  g_{1}g_{1}-2g_{2}\right)  =\frac{8}{\phi^{2}-as_{i}^{2}}\left[
-\frac{d}{d-2}\left(  \phi_{1}^{2}-as_{1i}^{2}\right)  +\left(  \phi\phi
_{2}-as_{i}s_{2i}\right)  \right]  , \label{trg2}%
\end{equation}%
\begin{equation}
\nabla_{\lambda}g_{1\mu}^{\lambda}-\partial_{\mu}g_{1\lambda}^{\lambda}%
=\frac{2}{\phi^{2}-as_{i}^{2}}\left[
\begin{array}
[c]{c}%
-\frac{1}{2a}\left(  \phi_{1}\partial_{\mu}\phi-as_{1i}\partial_{\mu}%
s_{i}\right) \\
+2\partial_{\mu}\left(  \phi\phi_{1}-as_{i}s_{1i}\right)  -g_{1\mu}^{\lambda
}\partial_{\lambda}\left(  \phi^{2}-as_{i}^{2}\right)
\end{array}
\right]  . \label{g1mn}%
\end{equation}
From the first of these we may solve algebraically for $Tr\left(
g_{2}\right)  ,$ and consider the second equation, along with (\ref{trg1}), as
equations of motion that restrict $g_{1\mu}^{\rho}$.

To show that there are solutions to the three prolongation equations
(\ref{trg1}-\ref{g1mn}), we provide an example with the following special
form, which of course is not the general case,%
\begin{equation}%
\begin{array}
[c]{ll}%
g_{1\mu}^{\nu}=A_{1}\left(  x\right)  \delta_{\mu}^{\nu},\;\;\; & g_{2\mu
}^{\nu}=A_{2}\left(  x\right)  \delta_{\mu}^{\nu},\\
\phi_{1}=B_{1}\left(  x\right)  \phi, & s_{1i}=B_{1}\left(  x\right)  s_{i},\\
\phi_{2}=B_{2}\left(  x\right)  \phi,\; & s_{2i}=B_{2}\left(  x\right)  s_{i},
\end{array}
\end{equation}
Furthermore, we use the Weyl gauge $\left(  \phi^{2}-as_{i}^{2}\right)
=\left(  2\kappa_{d}^{2}\right)  ^{-1}$ in Eq.(\ref{nonlinear}) to simplify
these equations. The three equations (\ref{trg1}-\ref{g1mn}) are then solved
by%
\begin{gather}
A_{1}\left(  x\right)  =\frac{2\kappa_{d}^{2}V\left(  \phi,s_{i}\right)
}{\left(  d-2\right)  }+c,\;\;\\
B_{1}\left(  x\right)  =-\frac{\kappa_{d}^{2}\left(  d-1\right)  }{2\left(
d-2\right)  }V\left(  \phi,s_{i}\right)  -\frac{1}{4}cd,\\
8B_{2}+2dA_{2}=d\left(  A_{1}^{2}+\frac{8B_{1}^{2}}{d-2}\right)  .
\end{gather}
where $c$ is an arbitrary constant. Hence the prolongations are determined by
the shadow fields, however one combination of $B_{2},A_{2}$ remains arbitrary.

Thus, we find that there are not sufficient equations to determine all of the
degrees of freedom $g_{1\mu}^{\nu},g_{2\mu}^{\nu},\phi_{1},\phi_{2}%
,s_{1i},s_{2i}$ that participated in the dynamics at $w=0.$ This is a sign
that there are gauge symmetries, so what cannot be determined by the equations
of motion must be a gauge degree of freedom, at least on-shell. We did
identify an off-shell gauge symmetry, namely the $\Lambda_{n}\left(  x\right)
$ in Eqs.(\ref{gaf1}-\ref{gag2}) which is sufficient to explain why one
function is gauge freedom in the example above, but the evidence is that there
is more gauge freedom. In fact more gauge symmetry should be expected as in
flat 2T-field theory \cite{2tstandardM}, where in the expansion in powers of
$w$ of \textit{matter fields} each coefficient except the zeroth order (i.e.
each prolongation) is a gauge degree of freedom. In flat 2T field theory the
prolongations decoupled completely from the shadow fields in flat space
\cite{2tstandardM} consistent with being gauge freedom. However, what we have
learned in this paper is that there also some that, rather than being gauge
freedom, are actually determined by the shadow fields via the geometry in
curved space $g_{1\mu}^{\nu},g_{2\mu}^{\nu}$ as seen in equations
(\ref{trg1}-\ref{g1mn}).

In any case, an outcome of our analysis is that there are non-trivial
prolongations which are determined by the shadow fields $\phi,s_{i},g_{\mu\nu
}$ up to gauge freedom. However, the shadow fields themselves $\phi
,s_{i},g_{\mu\nu}$ are determined self consistently by the action
(\ref{S0shad2}) only within the shadow, as in Eqs.(\ref{shadoweoms}),
independently of the prolongations.

\section{Concluding comments}

The decoupling of the dynamics of the shadow proven in the previous section is
significant because it shows that General Relativity in $d$ dimensions,
augmented with the Weyl symmetry, as expressed by the action (\ref{S0shad2}),
is the prediction of 2T-gravity for observers asking questions only in $d$
dimensions. Establishing this effective action principle, by analyzing the
equations of motion in detail as we did above, was one of the aims of our analysis.

This shows that the full physical (gauge invariant) information in $\left(
d+2\right)  $ dimensions is captured by the conformal shadow, so this is a
\textquotedblleft holographic\textquotedblright\ shadow. Turning this around,
we can also claim that usual General Relativity in $d$ dimensions, augmented
with the Weyl symmetry, is described directly in $d+2$ dimensions in the form
of 2T-gravity.

We have shown quite generally that the Weyl symmetry in 1T field theory is
directly related to higher spacetime general coordinate transformations that
include an extra time dimension. Therefore local Weyl symmetry is a strong
footprint of 2T-physics. Just like other gauge symmetries, there are
observable effects of the structure that this symmetry imposes on interactions.

As we have shown, as a consequence of 2T-gravity, the graviton and the scalars
must satisfy certain structures in 1T field theory. Dirac and Yang-Mills
fields can be included in a straightforward way except for inserting the
dilaton factors of $\phi^{\frac{2\left(  d-4\right)  }{d-2}}$ in Yang-Mills
kinetic terms and $\phi^{-\frac{d-4}{d-2}}$ in Yukawa terms (as in Table-1).
With these dilaton factors the crucial Weyl symmetry is intact in every
dimension $d.$ These are some of the footprints of 2T-gravity.

Some of the consequences of the emergent structures imposed by 2T-gravity were
outlined in the introduction and section (\ref{review}). Investigations of
physical effects in the context of cosmology and LHC physics are currently in
progress and will appear in future publications \cite{2tcosmology}.

\begin{acknowledgments}
We thank Yueh-Cheng Kuo, Guillaume Quelin and Bora Orcal at USC for helpful discussions
\end{acknowledgments}

\end{document}